\documentclass[manuscript]{aa}

\usepackage{graphicx}
\usepackage[varg]{txfonts}
\usepackage{hyperref}
\usepackage{xcolor}
\usepackage{soul}
\usepackage{multirow}
\usepackage{amsmath}
\usepackage{hyperref}
\usepackage{booktabs}
\usepackage{lipsum}
\usepackage{ulem}

\newcommand{\qradEQ}{\text{Q}_\text{rad}}
\newcommand{\qlossEQ}{\text{Q}_\text{loss}}

\newcommand{\qradint}{$ \int_t  \left< {\rm Q}_{\rm rad} \right> $}
\newcommand{\qlossint}{$\int_t \left< {\rm Q}_{\rm loss} \right> $}
\newcommand{\eintfluxint}{$\int_t \left<-\nabla \cdot ({\rm e}_{\text{int}} \mathbf{v})\right> $}

\newcommand{\peworkint}{$\int_t \left<-p \nabla \cdot \mathbf{v}\right>$}
\newcommand{\divqtcint}{$\int_t \left< \nabla \cdot \mathbf{q} \right>$}
\newcommand{\rhseintint}{$\int_t  \left< \partial {\rm e}_{\rm int}/\partial t \right>$ }

\usepackage{natbib}

\hypersetup{
    colorlinks=true,      
    linkcolor=blue,       
    citecolor=blue,       
    urlcolor=blue         
}

\newcommand{\serie}[2]{\multirow{#2}{*}{#1}}

\newcommand{\hd}[1]{\multirow{#1}{*}{RHD}}
\newcommand{\mhd}[1]{\multirow{#1}{*}{RMHD}}
\newcommand{\hp}[1]{\multirow{#1}{*}{Fixed}}
\newcommand{\ft}[1]{\multirow{#1}{*}{Dynamic}}

\newcommand{\zero}[1]{\multirow{#1}{*}{0}}
\newcommand{\mil}[1]{\multirow{#1}{*}{1000}}
\newcommand{\milqui}[1]{\multirow{#1}{*}{1500}}

\newcommand{\widthone}[1]{\multirow{#1}{*}{5.8}}

\newcommand{\mcols}[2]{\multicolumn{#1}{c|}{#2}}

\newcommand{\eitin}[1]{\multirow{#1}{*}{18.2}}
\newcommand{\cinqc}[1]{\multirow{#1}{*}{54.4}}

\newcommand{\bcc}[1]{\multirow{#1}{*}{Closed}}
\newcommand{\bgh}[1]{\multirow{#1}{*}{BHG}}
\newcommand{\bch}[1]{\multirow{#1}{*}{Charact.}}

\newcommand{\tini}{t$_\text{ini}$}

\newcommand{\zmax}{z$_\text{max}$}
\newcommand{\xmax}{x$_\text{max}$}

\usepackage{array}

\newcommand{\num}{\multirow{2}{*}{\#}}

\newcommand{\type}{\multirow{2}{*}{Type}}
\newcommand{\topbc}{\multirow{2}{*}{Top BC}}
\newcommand{\str}{\multirow{2}{*}{S}}

\usepackage{sidecap}

\usepackage{orcidlink}

\begin{document}

\title{Influence of numerical domain design in simulations of solar coronal heating with the MAGEC code}

\authorrunning{A. Navarro et al.}
\titlerunning{ A study of solar coronal heating with MAGEC}

\author{Anamar\'ia Navarro \and E. Khomenko \and T. Felipe \and S. J. Gonz\'alez Manrique}

\institute{Instituto de Astrof\'isica de Canarias, 38205 La Laguna, Tenerife, Spain 
\and
Dpto. de Astrof\'isica, Universidad de La Laguna, 38205 La Laguna, Tenerife, Spain 
\email{anavarro@iac.es} }

\date{\today}

\abstract{
Numerical simulations are essential for studying the complex structure of the solar atmosphere, especially in the context of the coronal heating problem. Understanding how numerical factors and boundary conditions (BC) influence the energetic structure of these models is crucial for unraveling the mechanisms behind the heating of the solar corona.
}
{
We investigate how the spatial dimensions of the computational domain, the upper BC and the vertical grid resolution influence the relaxed thermal structure, energetics, and long-term stability of simulations of the solar atmosphere in 2D with a vertical magnetic field.
}
{
Using the MAGEC code, we conducted several 2D parametric studies varying the domain width and height, vertical grid stretching, and top BC, among which we include the hot-plate, closed, unrestricted temperature, background heating configurations, and non-reflecting characteristics. We evaluated the resulting average temperature profiles to determine their parameter dependence, and temporal stability. Additionally, we analyzed the accumulated energetic budget as a function of height, isolating the contributions from radiative losses, thermal conduction, pressure work, and internal energy flux, allowing us to evaluate how the balance of these distinct energetic contributions shifts depending on the chosen top BC and explain the relaxed thermal profiles of the models.
}
{
The temperature structure of the models behaves significantly differently as a function of the domain height and top BC. We identified clear dependencies of these factors on the average temperature profile and the gradient of the horizontally averaged transition region (TR). In general, models with larger vertical domains showed higher average temperatures in the intermediate corona and exhibited steeper gradients across the average TR. Comparative analysis reveals that, among the less artificial BC, the non-reflecting characteristic are the most effective at delaying coronal cooling. Furthermore, we determined the threshold height at which simulations employing a dynamic top BC can effectively mimic the temporal behavior of fixed-temperature setups. We also established the maximum allowable degree of grid stretching in the corona that preserves physical accuracy while substantially reducing computational costs. Finally, the accumulated effects of the energetic terms changes significantly according to the chosen BC, with specific mechanisms acting as cooling agents in different layers of the atmosphere.
}
{
Our results demonstrate that in 2D the domain height, vertical grid design, and BC critically dictate the relaxed thermal structure and energetic balance of the simulated solar atmosphere.
}


\maketitle

\section{Introduction}\label{sec:intro}

This paper is a continuation of the work presented in \cite{magec}, in which the dynamics of both open and closed magnetic field configurations were investigated using realistic simulations of the solar atmosphere, extending from the sub-photospheric convective region to the corona with the MAGEC code. In that study, the open-field configuration exhibited significantly faster cooling than the closed-field case, raising questions about the influence of the upper boundary treatment on the thermal evolution of the system. The present work addresses these issues by examining in greater detail the effects of the temperature treatment in the top boundary conditions (BC) in regions with upward vertical magnetic field, as well as the impact of the domain height and grid structure on the thermal and dynamical response of the modeled solar atmosphere.

In the literature, there is no universal consensus on the most appropriate treatment of the upper BC in numerical models of the solar atmosphere. Consequently, a wide range of approaches has been developed, largely depending on the magnetic-field geometry, the extent of the computational domain, and whether the simulations are performed in 2D or 3D. In practice, boundary effects are often reduced by extending the numerical domain to the maximum height permitted by available computational resources, frequently in combination with an expansion of the grid spacing.

Let us name some of the main strategies frequently used in realistic 3D simulations in solar physics. In the works of \cite{2005ApJ...618.1020G} about the coronal heating through magnetic field line braiding, the temperature at the upper boundary was treated using a relaxation condition that gradually forced it toward the horizontally averaged temperature at the top of the computational domain. Their numerical domain extended up to 37 Mm above the solar surface and employed a vertical cell spacing varying from 150 to 250 km. \cite{2016A&A...585A...4C} simulated an enhanced-network magnetic configuration with an upper boundary transparent to outgoing perturbations through characteristic extrapolation of the MHD variables. In their setup, the numerical domain extended up to 14 Mm into the corona, with a vertical grid spacing ranging from 19 to 100 km.  \cite{2017ApJ...834...10R} evolved different atmospheric conditions from the quiet Sun to active regions using at the upper boundary a semi-transparent condition that damps vertical flows. Their numerical domain extended up to 41 Mm over the surface and employed a uniform vertical cell size of 64 km.  More recently, \cite{2022ApJ...937...91C} simulated quiet-Sun and large active-region atmospheres using an open BC, their domain height was located at 113 Mm in the corona, with a vertical grid of uniform 64 km resolution.

Maintaining coronal temperatures in 2D models is particularly challenging, especially in simulations with open or predominantly vertical magnetic-field configurations. This difficulty is evident in the simulations of \cite{2011ApJ...743..142H}, which considered several open-field setups. Their runs lasted 75 minutes, and in the absence of any imposed external heating, the corona cooled to approximately 250 kK. Their upper BC was transparent, based on characteristic extrapolation of the MHD variables. In their model, the top of the computational domain was located about 15 Mm above the surface, and the vertical grid spacing ranged from 28 to 150 km.  

In many cases, the temperature must be artificially maintained at the upper boundary to prevent excessive cooling, commonly denoted as the hot-plate condition. This approach was adopted by \cite{2011A&A...530A.124L}, who simulated a weakly magnetized region in 2D. The upper boundary was treated using characteristic methods, while the temperature was fixed at 250 kK. Their computational domain extended up to 14 Mm above the surface and employed a vertically stretched grid, with spacing ranging from 28 to 150 km. A similar strategy was adopted in the works of \cite{2015ApJ...812L..30I}, there the top boundary was open to vertical flows under a free-slip condition and the conductive flux was adjusted to maintain a prescribed coronal temperature. In that study, the magnetic field was vertical, the boxes extended up to 14 Mm and the vertical grid spacing was uniform, with a resolution of 32 km. A hot-plate condition was also employed by \cite{2016ApJ...822...18N}, who carried out a 2.5D simulation of flux emergence in a computational domain extended up to 14 Mm into the corona, with  vertical grid cells ranging from 19 to 90 km. Characteristic BC were applied, and the temperature at the upper boundary was fixed at 1 MK. 

However, in some cases when the magnetic field is not open or vertical, some 2D works have been able to sustain a hot corona without imposing a hot-plate condition at the top, this is the case of the works of \cite{2017ApJ...847...36M}, that simulated two medium-sized plage regions in a computational domain extended up to 40 Mm into the corona using a non-uniform vertical grid spacing ranging from 12 to 50 km. Alternatively, rapid cooling can be mitigated by introducing a background heating (BGH) term in the simulations (see, e.g., \cite{2022A&A...666A..28S}).

In this work, we aim to establish optimal numerical conditions for modeling a 2D solar atmosphere with an upward vertical magnetic field configuration, enabling the study of their relaxed thermal profile with minimal numerical artifacts introduced by the upper boundary. To achieve this, we carried out a comprehensive suite of experiments varying both the physical dimensions of the numerical boxes and the configuration of the BC.

Our analysis focuses specifically on the treatment of the top BC for the temperature, as well as on the choice of the minimum domain height required to obtain a relaxed thermal state that can hold high coronal temperatures without collapsing. To this end, our setups include traditional configurations with a fixed temperature at the top boundary—which mimic a maintained, heated corona and allow the system to reach a stationary equilibrium—as well as experiments with closed boundaries, non-reflecting characteristics, and a BGH. Additionally, we explore setups with a dynamic top boundary where the temperature is left unrestricted. These essentially function as coronal cooling experiments, allowing us to evaluate how long the models can remain thermally stable without imposing a hot-plate condition or adding extra heating. We investigate whether extending the numerical boxes upward can delay the inevitable cooling of the corona in these cases. To optimize this vertical extension, we also conducted experiments utilizing stretched grids, allowing us to determine the maximum extent to which the grid can be stretched and what are the implications on the simulations.

The paper is organized as follows. Section~\ref{sec:methodology} presents the overall methodology of the study. It begins with an introduction to the code and the numerical methods (Sect. \ref{subsec:numerical_methods}), followed by a description of the initial atmospheric configuration (Sect.\ref{subssec:ini_conditions}). The BC are detailed in Sect.\ref{subsec:bc}, while Sect.\ref{subsec:dimensions_and_resolution} discusses the domain dimensions, resolution, and grid setup. The BGH used in some of the simulations is presented in Sect.\ref{subsec:bgh}. The setup of the simulations is outlined in Sect.\ref{subsec:setup}.  The results are presented in Sect.~\ref{sec:results}. The influence of domain width is analyzed in Sect.\ref{subsec:box_width}, and the effects of domain height are examined in Sect.\ref{subsec:box_height}. The impact of grid stretching is explored in Sect.\ref{subsec:grid}. In Sect. \ref{subsec:time_stability_FT} we analyze the stability in time of the simulations without temperature restrictions in the top BC. The energetic balance of the simulations depending on their BC is discussed in Sect.\ref{subsec:energetic_analysis}. Finally, Sect. \ref{sec:conclusions} provides a summary of the main findings.

\begin{figure}
    \centering
    \includegraphics[height=0.18\textheight]{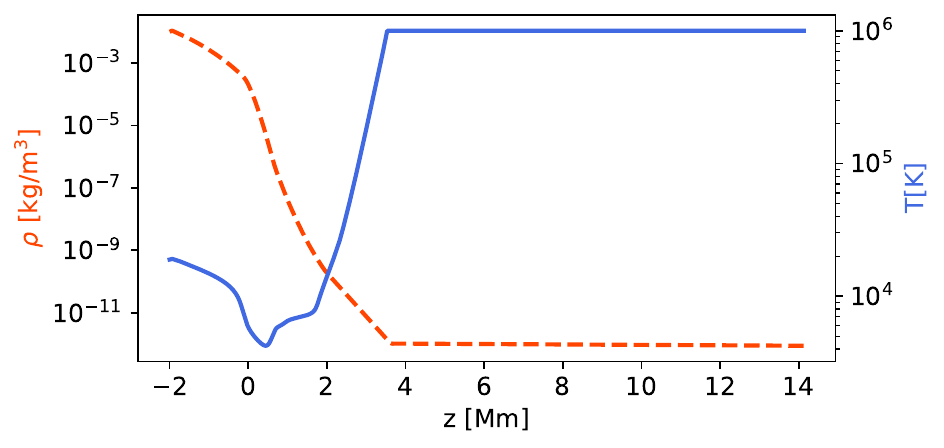}
    \caption{Initial vertical profiles of density and temperature for the simulation with the smallest domain height.}
    \label{fig:ini_profile}
\end{figure}

\section{Methodology \label{sec:methodology}}
\subsection{Numerical methods \label{subsec:numerical_methods}}

The simulations were performed with the MAGEC code \citep{magec}, a grid-based radiative-magnetohydrodynamic (RMHD) solver designed to model the dynamics of the solar atmosphere. The code accounts for thermal conduction and radiative energy exchange across different atmospheric layers. Radiative transfer is treated using an optically thick Local Thermodynamic Equilibrium (LTE) approximation in the lower regions (from the sub-surface convective zone to the chromosphere) and optically thin radiative losses in the upper chromosphere and corona. In the simulations presented in this paper, the treatment of radiative transfer is simplified by limiting it entirely to the grey approximation. A realistic equation of state (EOS) is employed up to the chromosphere, smoothly transitioning to an ideal EOS in the corona.

The numerical schemes used in this work are based on a finite-volume formulation with shock-capturing methods. Fluxes are computed using the HLLC Riemann solver \citep{hllc_mhd_Li} combined with a minmod slope limiter \citep{Roe1986} for spatial reconstruction, while time integration is performed with a third-order strong-stability-preserving Runge–Kutta scheme (SSP-RK3) \citep{gottlieb2001ssp} and the constrained transport method \citep{CT_evans} to limit the divergence of the magnetic field.

\subsection{Initial conditions\label{subssec:ini_conditions}}

The initial atmospheric structure of the simulations is constructed by combining several solar models. At heights $[-2,0]$ Mm, with $z$~=~0~Mm defined as the height where the continuum optical depth at 500 nm equals 1, we employed the solar convection zone model from \cite{1974SoPh...34..277S}. This setup is connected to the VAL-C model \citep{Vernazza_temperature_profile} used in the height range [0,~2]~Mm and is smoothly extended to $z$~=~3.5~Mm where the temperature stabilizes at 1MK. The transition between models is achieved by imposing the hydrostatic equilibrium and ensuring consistency with the EOS. Figure~\ref{fig:ini_profile} displays the initial density and temperature vertical profiles.  Random noise was introduced to the internal energy to initiate the convective instability.

\subsection{Boundary conditions \label{subsec:bc}}
Horizontal BC are periodic, while the bottom boundary is open to allow mass flow, with automated control applied to regulate the fluctuation of the mass and entropy, as done in \cite{2018A&A...618A..87K}. We apply different upper boundary configurations across our simulation sets. One set employs non-reflecting characteristic BC, utilizing the eigenvectors derived by \cite{roe_balsara_eigenvectors} and the formulation detailed by \cite{MAKAREMIESFARJANI2022105461}. The remaining cases impose the following conditions: symmetric conditions on the internal energy, density, $x$ and $y$ components of the velocity ($v_x$, $v_y$), and $z$ component of the magnetic field ($B_z$). Antisymmetric conditions are imposed on the $x$ and $y$ components of the magnetic field ($B_x$ and $B_y$). 

For the vertical velocity ($v_z$), in the case of the closed boundary simulations it is set to be antisymmetric, so it mimics a rigid wall. And for all the other ones we used an outflow condition, and it is set to zero when it is negative, thus preventing inflows:
\begin{align}    
    &{v_z}_{(N_z)} = \text{max}( {v_z}_{(N_z-2)} , 0) \, , \nonumber \\
    &{v_z}_{(N_z-1)} = \text{max}( {v_z}_{(N_z-2)}, 0) \, ,
\end{align}
where ${v_z}_{(N_z)}$ and ${v_z}_{(N_z-1)}$ are the top boundary points, with $N_z$ being the number of points in vertical direction. For simulations employing a hot-plate condition, the temperature at the top boundary is fixed at its initial value; we refer to these as fixed BC runs. Conversely, when the top temperature evolves dynamically based on symmetric conditions across the other variables, we designate these as dynamic BC runs.  
\begin{figure}
    \centering
    \includegraphics[width=0.5\textwidth]{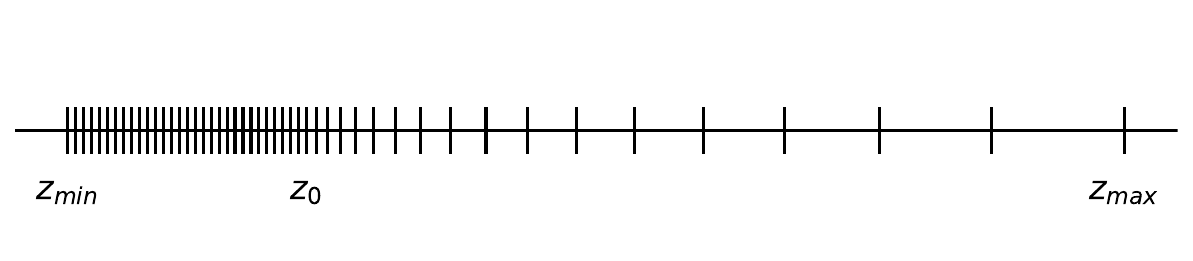}
    \caption{Vertical stretching of the non-uniform grid. The spacing is uniform between $z_\text{min}$ and $z_0$, and above $z_0$ it gradually increases with height. The stretching factor $S$, calculated using eq.~\eqref{eq:stretching_s}, is 2.8 for the grid shown here. This grid is for illustrative purposes.}
    \label{fig:non_uni_grid}
\end{figure}

\subsection{Dimensions and resolution \label{subsec:dimensions_and_resolution}}

The computational domains have widths of 5.8, 23 and 46.1~Mm and heights of  14.1, 18.2, 26.2, 34.3, 42.4, 46.4 and 54.4~Mm. In simulations with a uniform grid, the horizontal and vertical resolutions are 20 and 14~km, respectively. For simulations with a stretched vertical grid, the grid remains uniform up to $z$~=~10~Mm (with a 14 km spacing), above which the vertical spacing increases geometrically.
 
To calculate the stretching we construct the grid following the steps described in \cite{2007ApJS..170..228M}. 

In Fig. \ref{fig:non_uni_grid}, we present a sketch of the grid. The region between $z_\text{min}$ and $z_0$ has uniform spacing $dz_0$. After $z_0$ the spacing increases as a geometrical expansion of the stretching ratio $r$, which can be calculated as follows:
\begin{equation}
    dz_0 \left( r + r^2 + \dots + r^N \right) = z_\text{max} - z_0 \, , \label{eq:stretching}
\end{equation}
where $N$ is the number of points of the non-uniform section in the top. Equation \ref{eq:stretching} can be expressed in a closed form as:
\begin{equation}
   r\dfrac{1-r^N}{1-r} = \dfrac{z_\text{max} - z_0}{dz_0} \, .
\end{equation}
We solve the stretching ratio $r$ from this equation using the Newton-Raphson method. To quantify the degree of grid stretching, we define the parameter
\begin{equation}
S = \frac{N_z\text{(reference)}}{N_z}, \label{eq:stretching_s}
\end{equation}
where $N_z$ is the number of vertical grid points in the simulation, and $N_z\text{(reference)}$ corresponds to the number of vertical points in a simulation of reference with a uniform grid of vertical spacing $dz_0$.

\subsection{Background heating\label{subsec:bgh}}
A background heating (BGH) rate will be used in one simulation set. It is is given by
\begin{eqnarray}
Q_{\text{BGH}} = \frac{1}{2}  Q_0 \left[ 1 + \tanh\left(\frac{z - \tilde{z} }{w}\right) \right]\label{eq:Q_bgh}
\end{eqnarray}
where $Q_0 = 1.e^{-4}$ kg m$^{-1}$s$^{-3}$, $\tilde{z} = 3.5$ Mm and $w=0.1$ Mm. The simulations with this additional heating employ dynamic BC, and we refer to them as the BGH model.

\subsection{Setup\label{subsec:setup}}

\begin{table*}
\renewcommand{\arraystretch}{1.1}
\caption{Summary of simulation parameters.}
\centering
\begin{tabular}{|c|c|c|c|c|c|c|c|c|c|c|c|c|c|c|c|c|}
\cline{1-8} \cline{10-16}
\serie{Set}{2}&\num& \type    &\topbc   & \zmax   & \str             &  \tini    & \xmax         &\serie{}{37} &\serie{Set}{2} &\num& \type    &\topbc   & \zmax   & \tini     & \xmax         \\
              &    &          &         &  [Mm]   & \serie{-}{32}    & [s]       & [Mm]          &             &               &    &          &         &  [Mm]   & [s]       & [Mm]          \\ \cline{1-8} \cline{10-16}  
\serie{A}{4}  & 1  & \hd{2}   & \hp{11} &\eitin{4}&                  & \zero{2}  &  23.0         &             &\serie{G}{7}   & 1  & \hd{7}   &\bcc{14} &  14.1   &\zero{7}   & \widthone{41} \\ \cline{2-2}  \cline{8-8} \cline{11-11}  \cline{14-14}               
              & 2  &          &         &         &                  &           &  46.1         &             &               & 2  &          &         &  18.2   &           &               \\ \cline{2-2} \cline{3-3} \cline{7-7} \cline{8-8} \cline{11-11} \cline{14-14}       
              & 3  & \mhd{2}  &         &         &                  & \mil{2}   &  23.0         &             &               & 3  &          &         &  26.2   &           &               \\  \cline{2-2} \cline{8-8} \cline{11-11}  \cline{14-14}                
              & 4  &          &         &         &                  &           &  46.1         &             &               & 4  &          &         &  34.3   &           &               \\ \cline{1-1} \cline{2-2} \cline{3-3} \cline{5-5} \cline{7-7} \cline{8-8} \cline{11-11} \cline{14-14}  
\serie{B}{7}  & 1  & \hd{14}  &         &  14.1   &                  &\zero{14}  & \widthone{35} &             &               & 5  &          &         &  42.4   &           &               \\ \cline{2-2} \cline{5-5} \cline{11-11}  \cline{14-14}                     
              & 2  &          &         &  18.2   &                  &           &               &             &               & 6  &          &         &  46.4   &           &               \\ \cline{2-2} \cline{5-5} \cline{11-11}   \cline{14-14}                 
              & 3  &          &         &  26.2   &                  &           &               &             &               & 7  &          &         &  54.4   &           &               \\ \cline{10-10} \cline{2-2} \cline{5-5} \cline{11-11} \cline{12-12} \cline{14-14} \cline{15-15}         
              & 4  &          &         &  34.3   &                  &           &               &             &\serie{H}{7}   & 1  & \mhd{7}  &         &  14.1   &\mil{7}    &               \\ \cline{2-2} \cline{5-5} \cline{11-11} \cline{14-14}                      
              & 5  &          &         &  42.4   &                  &           &               &             &               & 2  &          &         &  18.2   &           &               \\ \cline{2-2}  \cline{5-5}  \cline{11-11} \cline{14-14}                  
              & 6  &          &         &  46.4   &                  &           &               &             &               & 3  &          &         &  26.2   &           &               \\  \cline{2-2}  \cline{5-5} \cline{11-11} \cline{14-14}                       
              & 7  &          &         &  54.4   &                  &           &               &             &               & 4  &          &         &  34.3   &           &               \\ \cline{1-1} \cline{2-2} \cline{4-4} \cline{5-5} \cline{11-11} \cline{14-14}      
\serie{C}{7}  & 1  &          & \ft{7}  &  14.1   &                  &           &               &             &               & 5  &          &         &  42.4   &           &               \\  \cline{2-2}  \cline{5-5}  \cline{11-11} \cline{14-14}               
              & 2  &          &         &  18.2   &                  &           &               &             &               & 6  &          &         &  46.4   &           &               \\ \cline{2-2}   \cline{5-5}  \cline{11-11} \cline{14-14}               
              & 3  &          &         &  26.2   &                  &           &               &             &               & 7  &          &         &  54.4   &           &               \\ \cline{10-10}\cline{2-2} \cline{5-5}  \cline{11-11} \cline{12-12} \cline{13-13} \cline{14-14} \cline{15-15}       
              & 4  &          &         &  34.3   &                  &           &               &             & \serie{I}{7}  & 1  & \hd{7}   &\bgh{14} &  14.1   &\zero{7}   &               \\ \cline{2-2} \cline{5-5} \cline{11-11} \cline{14-14}                  
              & 5  &          &         &  42.4   &                  &           &               &             &               & 2  &          &         &  18.2   &           &               \\  \cline{2-2}   \cline{5-5}  \cline{11-11}  \cline{14-14}             
              & 6  &          &         &  46.4   &                  &           &               &             &               & 3  &          &         &  26.2   &           &               \\  \cline{2-2}  \cline{5-5}   \cline{11-11} \cline{14-14}                         
              & 7  &          &         &  54.4   &                  &           &               &             &               & 4  &          &         &  34.3   &           &               \\ \cline{1-1}   \cline{2-2} \cline{3-3} \cline{4-4}\cline{5-5} \cline{7-7} \cline{11-11} \cline{14-14}
\serie{D}{7}  & 1  & \mhd{21} & \hp{7}  &  14.1   &                  &\mil{14}   &               &             &               & 5  &          &         &  42.4   &           &               \\  \cline{2-2}  \cline{5-5} \cline{11-11}   \cline{14-14}
              & 2  &          &         &  18.2   &                  &           &               &             &               & 6  &          &         &  46.4   &           &               \\   \cline{2-2}  \cline{5-5} \cline{11-11}  \cline{14-14}            
              & 3  &          &         &  26.2   &                  &           &               &             &               & 7  &          &         &  54.4   &           &               \\ \cline{10-10}  \cline{2-2} \cline{5-5} \cline{11-11} \cline{12-12} \cline{14-14} \cline{15-15}  
              & 4  &          &         &  34.3   &                  &           &               &             & \serie{J}{7}  & 1  & \mhd{7}  &         &  14.1   &\mil{7}    &               \\  \cline{2-2}  \cline{5-5} \cline{11-11} \cline{14-14}                 
              & 5  &          &         &  42.4   &                  &           &               &             &               & 2  &          &         &  18.2   &           &               \\   \cline{2-2} \cline{5-5} \cline{11-11} \cline{14-14}                 
              & 6  &          &         &  46.4   &                  &           &               &             &               & 3  &          &         &  26.2   &           &               \\  \cline{2-2} \cline{5-5} \cline{11-11} \cline{14-14}                 
              & 7  &          &         &  54.4   &                  &           &               &             &               & 4  &          &         &  34.3   &           &               \\ \cline{1-1}  \cline{2-2} \cline{4-4} \cline{5-5}\cline{11-11} \cline{14-14}                         
\serie{E}{7}  & 1  &          &\ft{14}  &   14.1   &                  &           &               &             &               & 5  &         &         &  42.4   &           &               \\   \cline{2-2} \cline{5-5}  \cline{11-11} \cline{14-14}                                            
              & 2  &          &         &  18.2   &                  &           &               &             &               & 6  &          &         &  46.4   &           &               \\ \cline{2-2}    \cline{5-5} \cline{11-11} \cline{14-14}                                            
              & 3  &          &         &  26.2   &                  &           &               &             &               & 7  &          &         &  54.4   &           &               \\ \cline{10-10} \cline{2-2} \cline{5-5}\cline{11-11} \cline{12-12} \cline{13-13} \cline{14-14} \cline{15-15}       
              & 4  &          &         &  34.3   &                  &           &               &             & \serie{K}{7}  & 1  & \hd{7}   &\bch{14} &  14.1   &\zero{7}   &               \\  \cline{2-2}   \cline{5-5} \cline{11-11}  \cline{14-14}                       
              & 5  &          &         &  42.4   &                  &           &               &             &               & 2  &          &         &  18.2   &           &               \\  \cline{2-2}   \cline{5-5} \cline{11-11}  \cline{14-14}                              
              & 6  &          &         &  46.4   &                  &           &               &             &               & 3  &          &         &  26.2   &           &               \\  \cline{2-2}    \cline{5-5}  \cline{11-11} \cline{14-14}             
              & 7  &          &         &  54.4   &                  &           &               &             &               & 4  &          &         &  34.3   &           &               \\ \cline{1-1}\cline{2-2} \cline{5-5}  \cline{6-6}  \cline{7-7} \cline{11-11} \cline{14-14} 
\serie{F}{7}  & 1  &          &         &\cinqc{7}& 1.0              &\milqui{7} &               &             &               & 5  &          &         &  42.4   &           &               \\   \cline{2-2}  \cline{6-6} \cline{11-11} \cline{14-14}                
              & 2  &          &         &         & 2.2              &           &               &             &               & 6  &          &         &  46.4   &           &               \\   \cline{2-2}  \cline{6-6} \cline{11-11} \cline{14-14}                          
              & 3  &          &         &         & 2.7              &           &               &             &               & 7  &          &         &  54.4   &           &               \\  \cline{10-10} \cline{2-2}\cline{6-6} \cline{11-11} \cline{14-14}  \cline{15-15}                               
              & 4  &          &         &         & 3.4              &           &               &             & \serie{L}{7}  & 1  & \mhd{7}  &         &  14.1   &\mil{7}    &               \\   \cline{2-2}  \cline{6-6} \cline{11-11}  \cline{12-12} \cline{14-14}                    
              & 5  &          &         &         & 3.5              &           &               &             &               & 2  &          &         &  18.2   &           &               \\  \cline{2-2}  \cline{6-6} \cline{11-11} \cline{14-14}                           
              & 6  &          &         &         & 3.7              &           &               &             &               & 3  &          &         &  26.2   &           &               \\  \cline{2-2}  \cline{6-6} \cline{11-11} \cline{14-14}                               
              & 7  &          &         &         & 4.1              &           &               &             &               & 4  &          &         &  34.3   &           &               \\ \cline{1-8}  \cline{6-6}  \cline{8-8}  \cline{11-11} \cline{14-14}     
\mcols{9}{ }                                                                                                   &               & 5  &          &         &  42.4   &           &               \\ \cline{11-11} \cline{14-14}                          
\mcols{9}{ }                                                                                                   &               & 6  &          &         &  46.4   &           &               \\ \cline{11-11} \cline{14-14}                             
\mcols{9}{ }                                                                                                   &               & 7  &          &         &  54.4   &           &               \\  \cline{10-16}  \cline{14-14}                        
\end{tabular}  
\label{tab:description}
\tablefoot{The column Type indicates whether the simulation is RHD or RMHD; Top BC specifies the top BC for the temperature (fixed, dynamic, closed, non-reflecting characteristic or BGH model); $z_\mathrm{max}$ denotes the top-domain height; $S$ is the grid-stretching factor (Eq.~\ref{eq:stretching_s}); $t_\mathrm{ini}$ gives the initial simulation time; and $x_\mathrm{max}$ indicates the horizontal domain width. }
\end{table*}

A total of 81 simulations were performed, covering most of the combinations of the parameters. The main properties of each model are summarized in Table~\ref{tab:description}, where we label each type with letter from A to L. The column Type indicates whether the simulation is RHD (radiative-hydrodynamic) or RMHD; Top BC specifies the top BC for the temperature (fixed, dynamic, closed, non-reflecting characteristic) or the model BGH; $z_\mathrm{max}$ denotes the top-domain height; $S$ is the grid-stretching factor (Eq.~\ref{eq:stretching_s}); $t_\mathrm{ini}$ gives the initial simulation time; and $x_\mathrm{max}$ indicates the horizontal domain width.

All simulations were initially evolved in a purely RHD regime with different domain heights, widths, and top BC for temperature (fixed, dynamic, closed or with BGH). From each of these runs, a snapshot at $t = 1000$~s was selected, and a uniform vertical magnetic field of 10~G was introduced. After the magnetic field was imposed, the simulations were continued in the full RMHD regime for 2.7 hours of solar time. 

Specifically, the base snapshot for simulation A3 was taken from the RHD simulation A1, and the one for A4 from A2. For simulations in set D, the snapshots were taken from the corresponding simulations in set B, while for set E they were taken from the set C. The snapshots for the stretched-grid simulations in set F were taken from simulation B7 and the top BC were changed to dynamic. Similarly, the base snapshots for sets H, J, and L, were extracted from the corresponding simulations in sets G, I, and K, respectively.

In the RHD simulations, the only non-ideal term considered is the radiative loss in the optically thick sub-photospheric layers. In the RMHD simulations, both the radiative losses in the optically thick and optically thin regions are included, together with thermal conduction. While the purely RHD setup does not represent a realistic solar atmosphere, this deliberate separation of physical ingredients enables a controlled comparison between models that exclude and include coronal processes, allowing us to examine their contribution to the thermal and dynamical response of the atmosphere.

The simulations presented in this work are, in general, very similar to those described in \cite{magec} for the configuration with a vertical magnetic field. In that study, a fixed temperature was imposed at the top boundary, and the computational domain had a width of 46.1~Mm. The grid was uniform and had the same spatial resolution as the uniform-grid simulations in the present work. We refer the reader to that paper for detailed visualizations of the temperature, density, magnetic field, and velocity fields, as well as for online videos illustrating their temporal evolution.

\section{Results \label{sec:results}}
\subsection{Dependence on the simulation box width\label{subsec:box_width}}
\begin{figure}
    \centering
    \includegraphics[width=0.6\linewidth]{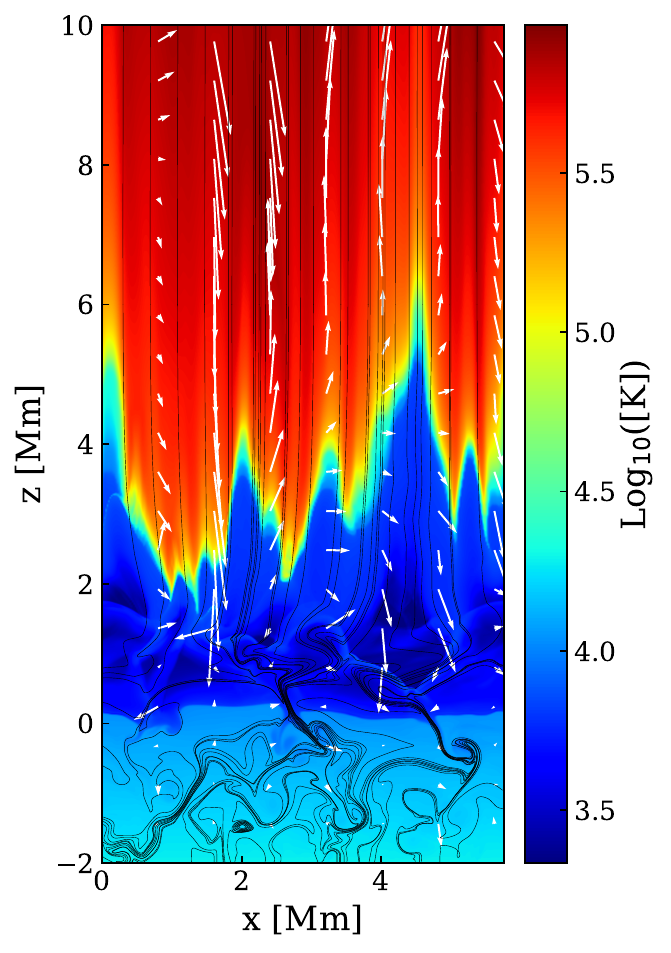}
    \caption{Colormap of the logarithmic temperature distribution at t = 2000~s for the RMHD simulation with the fixed top BC (run B2). The white arrows indicate the velocity field, with lengths proportional to the local flow speed and the black lines are the magnetic field lines. }
    \label{fig:te_colormap}
\end{figure}
\begin{figure}[h]
    \centering
    \includegraphics[width=0.49\textwidth]{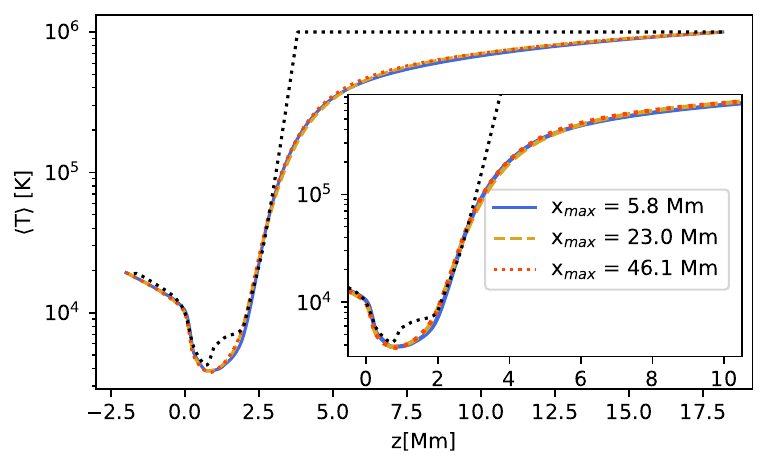}
    \caption{Horizontally and temporally averaged temperature profiles from the RMHD simulations with fixed top BC and domain widths of 5.8~Mm, 23.0~Mm, and 46.1~Mm, corresponding to simulations B2, A3, and A4, respectively. The black dotted line shows the initial temperature profile. The temporal average covers the interval from $t$~=~2400~s to 3000~s.}
    \label{fig:te_average_width}
\end{figure}
The overall structure of the simulations is illustrated in Fig. \ref{fig:te_colormap} by the colormap of the logarithmic temperature for the run with the smallest horizontal width (5.8~Mm, simulation B2), displayed up to $z$~=~10~Mm. The snapshot corresponds to an intermediate stage at $t$~=~2000~s. The figure shows the formation of three convective cells with characteristic sizes of about 2~Mm. As expected from the physical grounds, the flows in the upper chromosphere and corona are predominantly vertical, following the magnetic field lines, while below the surface ($z$~=~0), they decouple and the magnetic field is twisted due to the convective motions.

The choice of the domain width for the parametric study was examined by comparing this simulation with two additional runs using larger horizontal extents of $x_\text{max}$~=~23.0 and 46.1~Mm (simulations A3 and A4, respectively). Figure~\ref{fig:te_average_width} shows the horizontally and temporally averaged temperature profiles for these three runs as a function of height. The profiles are very similar. The difference quantified by the relative error, 
\begin{equation}
\text{relative error} = \frac{T_2-T_1}{T_1}100\% \, ,  \label{eq:relative_error} 
\end{equation}
is approximately 20\%; however, differences exceeding 10\% are confined to only 5\% of the domain around $z \approx$ 2.4 Mm. Therefore, we consider the box width of 5.8 Mm to be an appropriate size for our parametric study. We clarify that the reduced horizontal extent of our domain is specific to the highly vertical magnetic field configuration chosen for these simulations, where the plasma dynamics are predominantly field-aligned in the corona.

For this averaged temperature profile, as well as for all others presented in this paper, we use the time interval $t$~=~2400–3000~s. The black dotted line represents the initial temperature profile. In some cases, such as this one, a zoomed-in view of the lower part of the domain (up to $z$~=~10~Mm) is also included.

\subsection{Dependence on the simulation box height\label{subsec:box_height}}

\begin{figure*}
    \centering
   \includegraphics[width=0.98\textwidth]{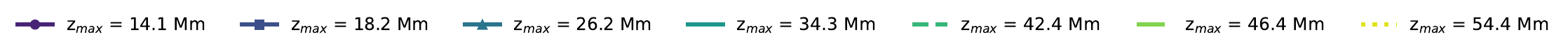}\\
    \includegraphics[width=0.31\textwidth]{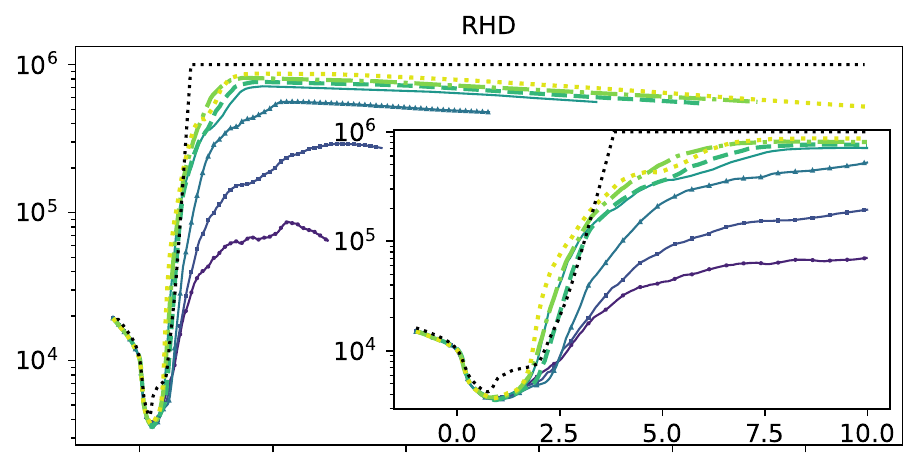}
    \includegraphics[width=0.31\textwidth]{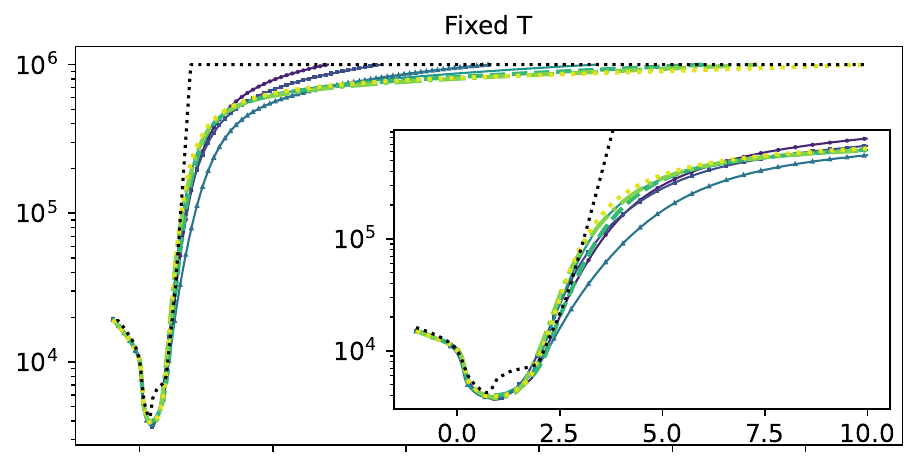}
    \includegraphics[width=0.31\textwidth]{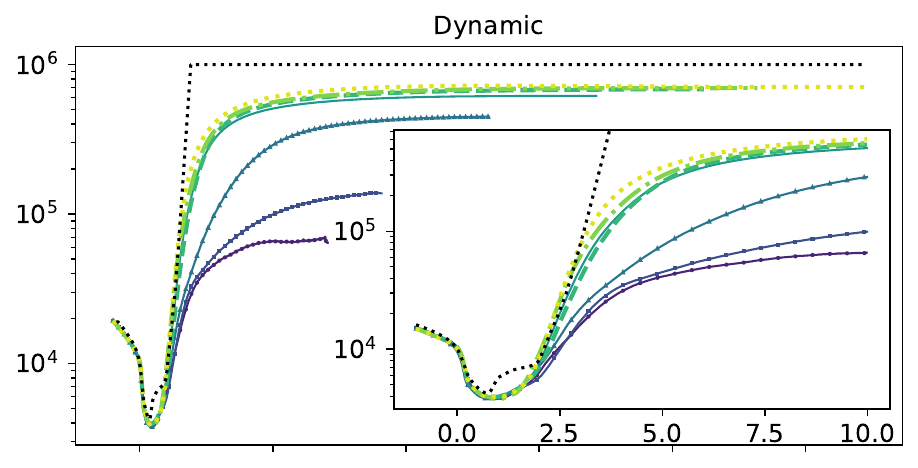}\\    
    \includegraphics[width=0.31\textwidth]{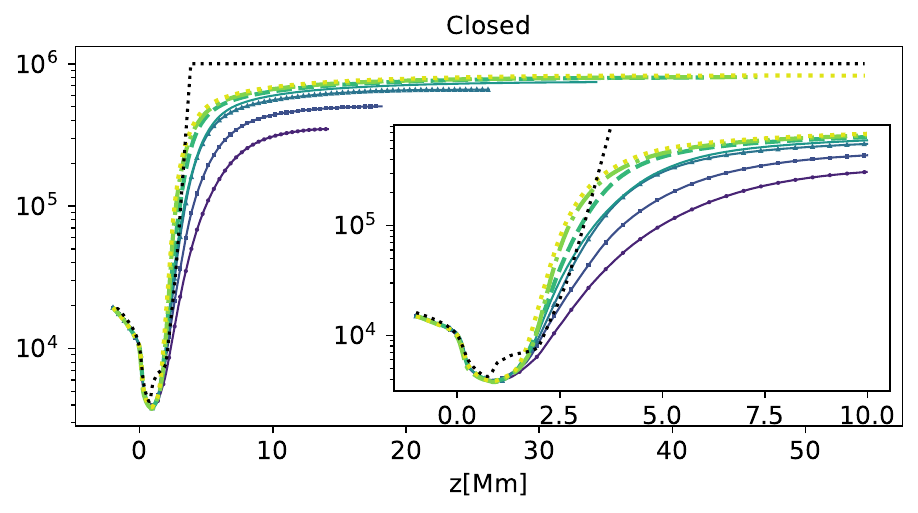}    
    \includegraphics[width=0.31\textwidth]{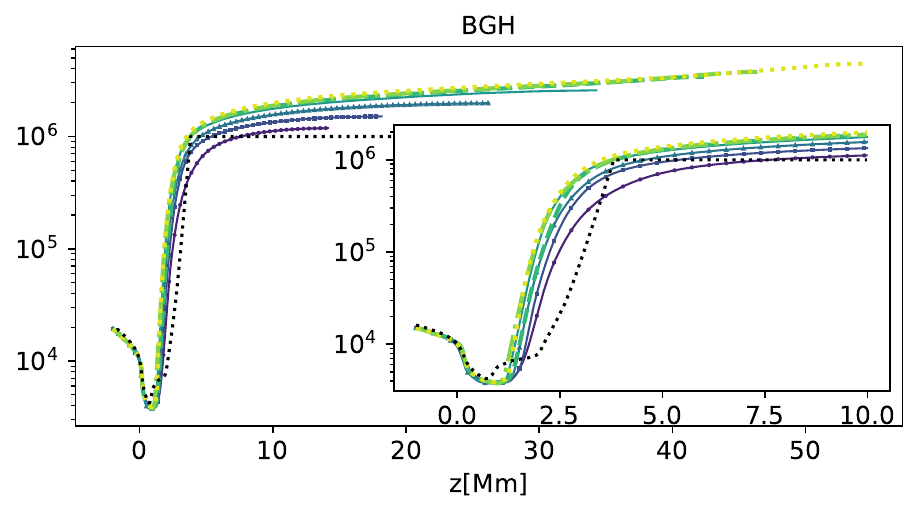}
    \includegraphics[width=0.31\textwidth]{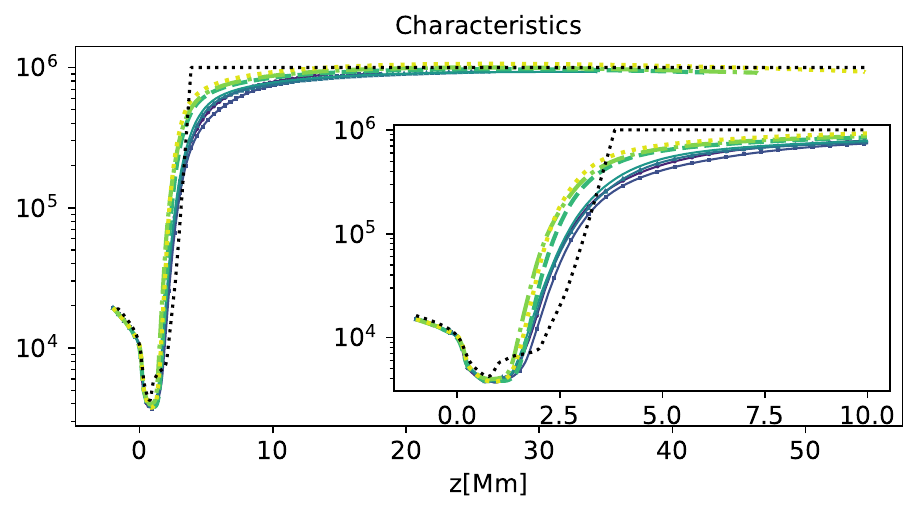}
\caption{Horizontally and temporally averaged temperature profiles. Panels from top left to right bottom display simulations with varying domain heights across five sets: RHD with dynamic top BC (set C), RMHD with fixed top BC (set D), RMHD with dynamic top BC (set E), RMHD with closed BC (set H), RMHD with BGH (set J) and RMHD with non-reflecting characteristic BC (set L). The black dotted line marks the initial temperature profile. Insets provide a zoomed-in view. The temporal average spans $t = 2400$--$3000$~s. The units are in kelvin.}
    \label{fig:te_average_heights}
\end{figure*}

\begin{figure*}[!ht]
    \centering
\includegraphics[width=0.6\textwidth]{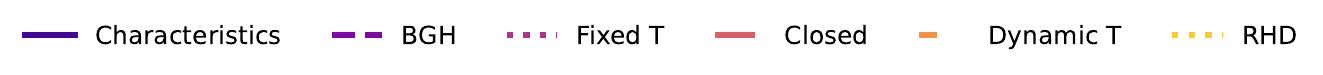}\\
    \includegraphics[width=0.3\textwidth]{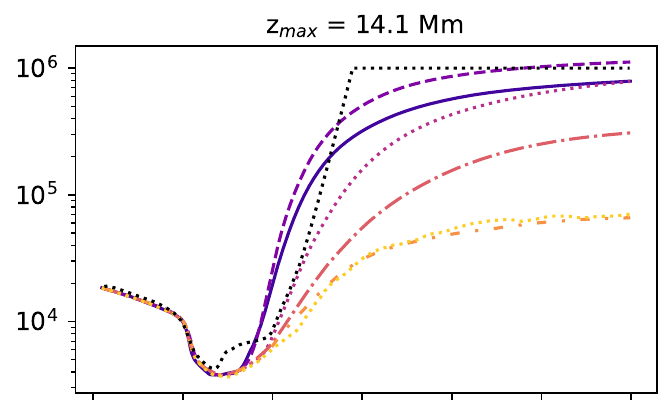}
     \includegraphics[width=0.3\textwidth]{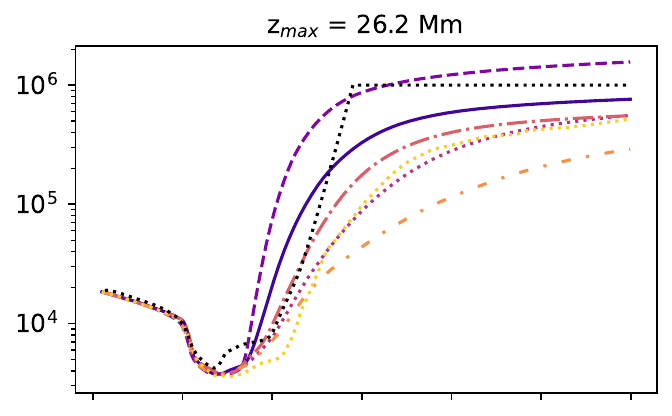}\\
     \includegraphics[width=0.3\textwidth]{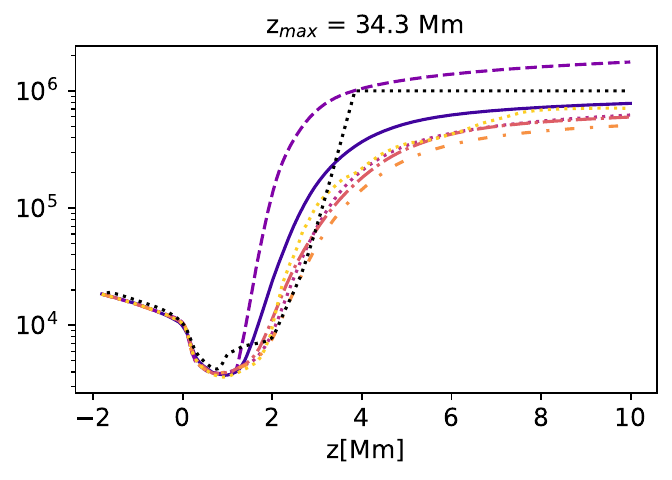}
     \includegraphics[width=0.3\textwidth]{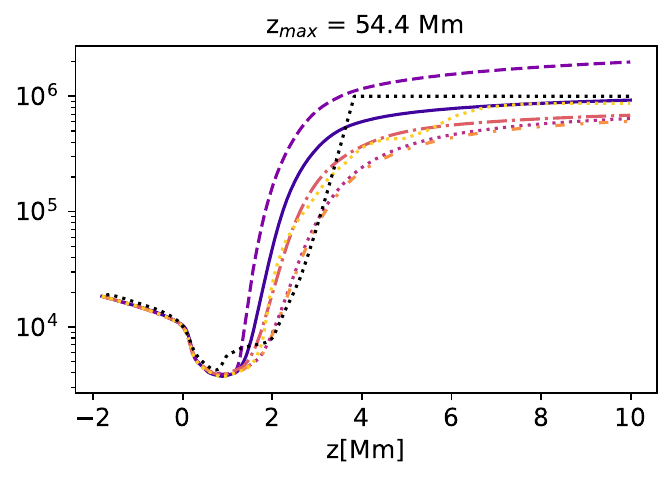}   
    \caption{Horizontally and temporally averaged temperatures of some of the simulations presented in \ref{fig:te_average_heights} grouped by box heights. The black dotted line represents the initial temperature profile. The temporal average covers the interval from $t$~=~2400~s to $t$~=~3000~s.  The profiles are shown from the bottom boundary up to 10 Mm. The units are in kelvin.}
    \label{fig:comparisons_FT_vs_HP}
\end{figure*}

We begin our parametric study by comparing the horizontally and temporally averaged temperature profiles for most of the sets. Figure~\ref{fig:te_average_heights} summarizes these results. From left top to right bottom, the panels display: The RHD simulations with dynamic top BC (set~C), the RMHD simulations with fixed top BC (set~D), the RMHD simulations with dynamic top BC (set~E), the RMHD simulations with closed top BC (set~H), the RMHD BGH simulations (set~J) and the RMHD simulations with non-reflecting characteristic BC.

Across all the sets a general trend can be identified: none of the models recover the chromospheric temperature observed in the initial atmosphere, most likely due to insufficient spatial resolution, the assumption of local thermodynamic equilibrium (LTE), instantaneous ionization balance, and the absence of a proper treatment of partially ionized plasma effects in the chromosphere. 

We will compare the general structure of the profiles, the temperatures in the lower corona, and the sharpness of the vertical confinement of the dynamic chromosphere-corona interface by describing its steepness; for simplicity, we refer to this as the average transition region (TR) steepness of the models. However, we emphasize that this does not refer to the intrinsic thickness of the local TR layer.

In the RHD simulations with dynamic top BC (top left panel, set~C), for simulations with domain heights above 26~Mm, the temperature peaks just above their respective average TR, followed by a gradual decrease with height. The simulation with a height of 34.3~Mm maintains the initial temperature profile up to $z$~=~3~Mm. Taller runs exhibit a sharper average TR than the initial profile. At coronal heights like $z$~=~10~Mm, these taller simulations retain higher temperatures, with the last four runs nearly recovering the initial profile at this height before decaying above it. 

In the RMHD simulations with fixed top BC (top right panel, set~D), except for the two shortest runs, the temperature profiles show a consistent trend with increasing domain height: taller boxes hold hotter temperatures in the low coronae and sharper average TR. These simulations also recover the initial temperature profile up to $z$~=~3~Mm, except for the chromospheric heights. 

In the RMHD simulations with dynamic top BC (middle left panel, set~E), the tallest runs sustain a hotter corona, with the maximum temperature located near the top of the computational domain. The three shortest simulations deviate strongly from the initial profile, whereas the four tallest ones recover the initial structure up to $z\sim3$~Mm, similar to the behavior observed in set~D. 

In the RMHD simulations with a closed top BC (middle right panel, set~H), the temperature profiles are generally slightly hotter above the TR and exhibit sharper average TR compared to the previous case. Furthermore, the height dependence of the boxes among these runs is significantly less pronounced, with the profiles clustering closer together. Except for the two shortest models, these simulations feature an average TR that is sharper than the initial state (black dotted line); for the tallest simulations, this enhanced gradients region spans from $\sim 2\text{ Mm}$ to $\sim 3.5\text{ Mm}$.

The RMHD simulations including BGH (bottom left panel, set~J; see Eq. \eqref{eq:Q_bgh} exhibit highly abrupt average TR at $z \sim$ 2 Mm. Compared to the initial profile, the average TR is shifted upward from its original position at $\sim$ 1 Mm and is notably steeper due to the effects of the BGH. Consistent with previous simulation sets, the steepness of the average TR increases with the box height. Additionally, the coronal temperature is also height-dependent, where taller domains sustain a hotter corona.

The RMHD simulations with non-reflecting characteristic BC (bottom right panel, set L) exhibit a steep average TR similar to the models with BGH. In these runs, the sensitivity of the average temperature profiles to box height is less pronounced than in previous cases; notably, the temperature values at 10 Mm remain nearly identical across all domain heights. The profiles split into two distinct regimes: the four shortest models exhibit nearly identical behavior, while the three tallest boxes form a separate, cohesive group.

The models employing closed boundaries, BGH heating, and non-reflecting characteristic BC exhibit clear, monotonic behavior as a function of domain height. In contrast, the RHD runs and those utilizing dynamic top BC display a more complex trend. This discrepancy is likely due to the half-open nature of the upper boundary. Although the BGH models share this half-open BC, their dynamics are heavily dominated by the explicit volumetric heating. Meanwhile, the closed boundary runs completely prohibit mass and energy fluxes, and the non-reflecting characteristic BC handles the no-inflow constraint far more smoothly.

Having analyzed the individual simulation sets, we now contrast the runs across all previous configurations to evaluate the impact of the domain height. This cross-set comparison is presented in Fig.~\ref{fig:comparisons_FT_vs_HP}, which displays the horizontally and temporally averaged temperature profiles grouped into four panels, each corresponding to a specific vertical extent. By structuring the figure this way, simulations from different sets but with identical domain heights can be directly compared. For a detailed view of the lower atmosphere, these profiles are restricted from the bottom boundary up to $10\text{ Mm}$.

Across all four panels of Fig.~\ref{fig:comparisons_FT_vs_HP}, the temperature profiles exhibit a consistent behavior in the lower layers ($z <$ 1 Mm), where the photosphere and the minimum temperature region remain largely decoupled from the different physics of the models, the BC, and the domain height, showing almost identical overlapping profiles. However, above $z \sim$ 1 Mm, the profiles diverge significantly depending on the specific simulation set. The most prominent trend is the behavior of the BGH models (solid dark blue line): across all domain heights, it consistently produces the steepest, most abrupt average TR, which is systematically shifted downward as previously seen in Fig.~\ref{fig:te_average_heights}. It is also the only model that successfully sustains temperatures above 1 MK within the first 10 Mm of the atmosphere, regardless of the domain height. Following the BGH models, the setups with non-reflecting characteristic BC produce the next steepest TR transition and consistently maintain the second-highest temperature profile in the corona.

\begin{figure*}[!ht]
    \centering
    \includegraphics[width=0.6\textwidth]{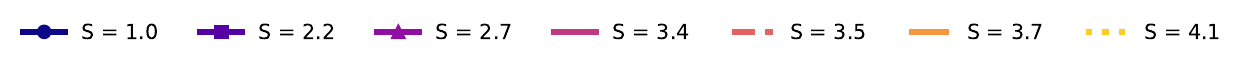}\\
    \includegraphics[width=0.4\textwidth]{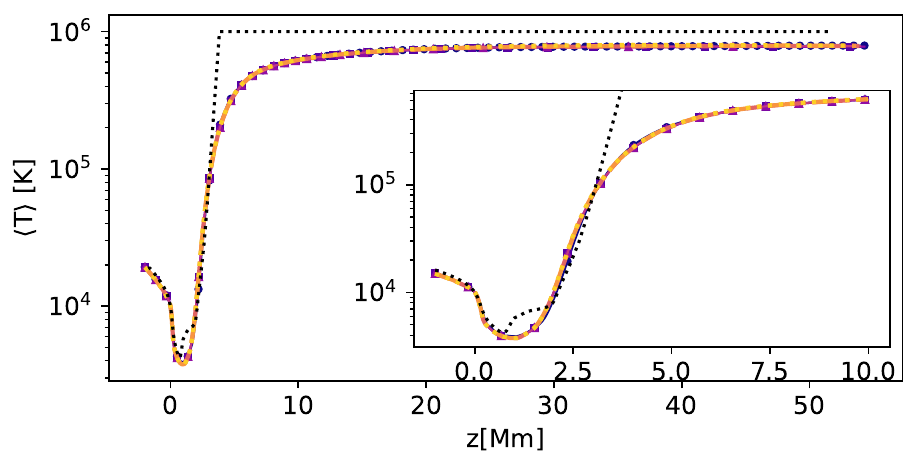}    
    \includegraphics[width=0.4\textwidth]{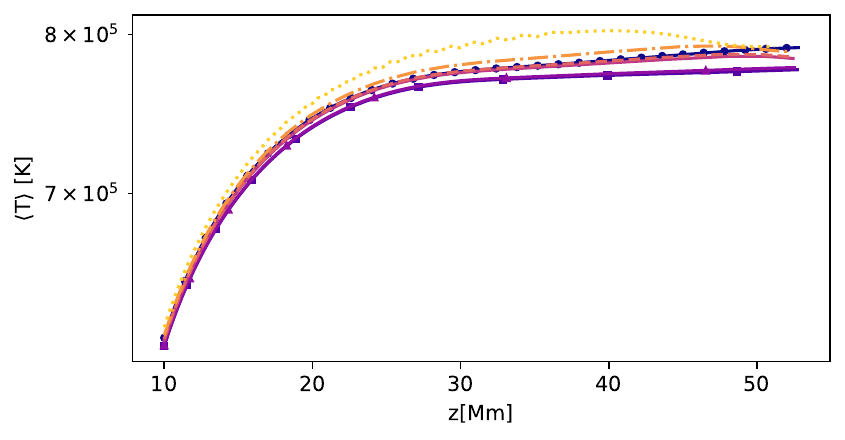}\\
    \includegraphics[width=0.4\textwidth]{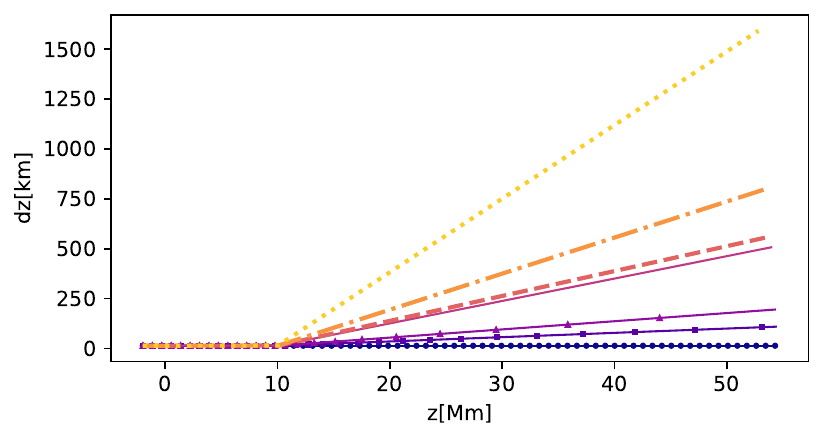}
    \includegraphics[width=0.4\textwidth]{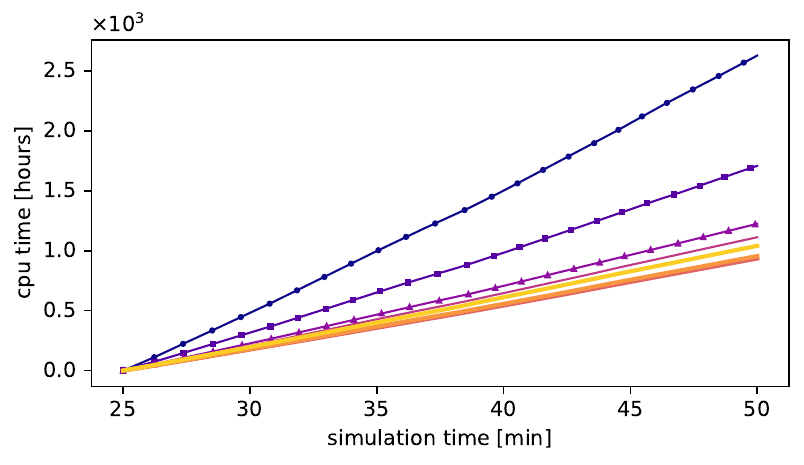}
\caption{Experiments with stretched grids (set F) at different stretching levels (S). Top-left panel: horizontally and temporally averaged temperature; zoomed-in view from the bottom up to z~=~10~Mm. Top-right panel: zoomed-in averaged temperature from z~=~10~Mm to the top of the domain. Bottom-left panel: grid spacing $dz$ versus height. Bottom-right panel: CPU time as a function of simulation time for each case. The temporal average covers the interval from $t$~=~2400~s to $t$~=~3000~s. } \label{fig:plots_stretched_grid}
\end{figure*}

With the exception of the BGH and fixed-temperature BC models, which maintain a sustained corona, these runs represent cooling experiments where thermal retention in the low corona ($\sim 6$~Mm) depends strongly on domain height. As $z_{\max}$ increases, thermal decay slows substantially, causing the temperature profiles of these models to become increasingly similar, cluster together, and remain at significantly higher values in taller computational domains.

When comparing only the runs with fixed and dynamic top BC we have that for heights below $z = 10~\mathrm{Mm}$, the relative error (Eq. \ref{eq:relative_error}) between each pair of runs is as follows: $47\%$ for a box height of $34.3~\mathrm{Mm}$, $37\%$ for $42.4~\mathrm{Mm}$, $24\%$ for $46.4~\mathrm{Mm}$, and $8\%$ for $54.4~\mathrm{Mm}$. Although the difference is not uniform, its maximum is localized at $z \approx 2.5~\mathrm{Mm}$. From these experiments, we found that simulations with top domain heights larger than $34.3~\mathrm{Mm}$ produce an average atmospheric temperature that is increasingly similar to that of runs with a fixed temperature at the top boundary. This agreement holds for at least 30 minutes of simulated solar time, between $t = 1200~\mathrm{s}$ and $3000~\mathrm{s}$. The stability over longer times will be discussed in Sect. \ref{subsec:time_stability_FT}.

\subsection{Dependence on the grid stretching \label{subsec:grid}}

From the previous experiments with dynamic BC, we found that simulations with box heights greater than 34.3 Mm produce more stable results and resemble the hot-plate simulations. Our tallest simulation, with a height of 54.4 Mm, produced an almost identical temperature profile to that of the hot-plate case, at least for heights below $z$~=~10~Mm. This experiment used 4032 grid points in the vertical direction. Running simulations with such a large number of points is computationally expensive, even in 2D. To reduce the computational cost while maintaining accuracy, we explored the use of a stretched grid in the simulations with dynamic top BC. For this purpose, we used as a baseline the snapshot at $t=1500$ s from the hot-plate simulation with the tallest box (run D7), switched the top BC to dynamic, and performed a parametric study (Set F) with different levels of grid stretching. 

In all cases, the grid stretching starts from $z_0$~=~10~Mm upwards. The reference simulation used to calculate the stretching level $S$ in Eq. \ref{eq:stretching_s} has 4032 grid points in the vertical direction. This parametric study considers stretching factors $S=1.0$ (the uniform-grid reference case), 2.2, 2.7, 3.4, 3.5, 3.7, and 4.1, which correspond to simulations with 4032, 1812, 1500, 1176, 1152, 1080, and 984 vertical grid points, respectively.

The bottom-left panel of Fig. \ref{fig:plots_stretched_grid} shows the vertical grid spacing ($dz$) for the different simulations. The smallest spacing corresponds to the uniform grid, with a minimum value of 14 km, and increases up to about 1600 km for the most stretched configuration. The top-left panel presents the horizontally and temporally averaged temperature over the time interval from $t$~=~2400~s to $t$~=~3000~s. As usual, the black dotted line indicates the initial temperature profile. The differences among the profiles are almost imperceptible; to better visualize them, we include a zoomed-in view up to $z$~=~10~Mm. This plot shows that the average temperature profiles are very similar, nearly identical.

To further investigate the differences, the top-right panel shows a zoomed-in view of the upper region, from $z$~=~10~Mm upward. In this plot, small deviations become visible, particularly in the most stretched grid, which differs slightly from the others between approximately $z$~=~20~Mm and the top boundary. However the maximum difference does not go beyond 1.5kK or a 2\% with respect to the profile of the reference simulation. 

Finally, the bottom right-panel compares the CPU time (in hours) among the different runs. As expected, simulations with more stretched grids require less computational time, achieving a speed-up of up to a factor of 2.81 in the best case. The most stretched case was among the fastest runs, though not the fastest one. This is likely due to the fact that, as seen in the temperature plots, this simulation becomes somewhat noisy in the coronal region as a consequence of the high level of stretching.

\subsection{Stability in time} \label{subsec:time_stability_FT}

\begin{figure}
    \centering
    \includegraphics[width=0.8\linewidth]{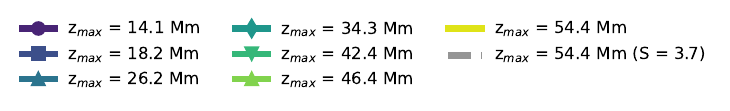}\\
    \includegraphics[width=0.8\linewidth]{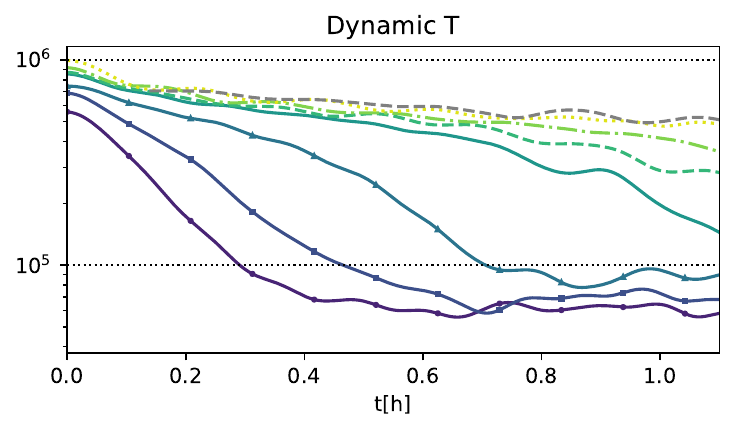}
    \includegraphics[width=0.8\linewidth]{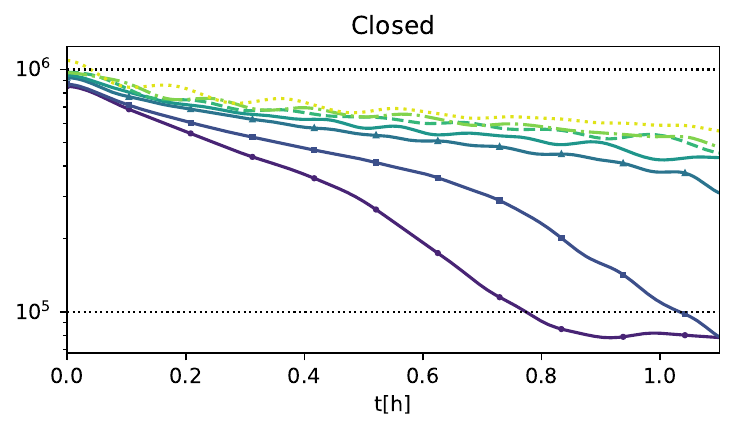}
    \includegraphics[width=0.8\linewidth]{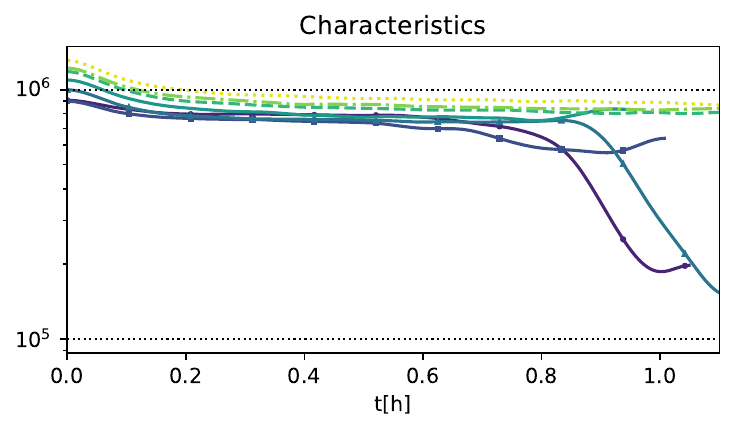}
    \caption{Average temperature at $z = 10~\mathrm{Mm}$ as a function of time. Top panel corresponds to RMHD simulations with dynamic BC (set E) and one run with stretched grid ($S$=3.7) from set F. Middle panel for simulations with closed BC (set H) and bottom panel for simulations with non-reflecting characteristic BC (set L). The units are in kelvin. }
    \label{fig:average_longime}
\end{figure}

To assess the long-term stability of the simulations under dynamic, closed, and non-reflecting top BC, Fig. ~\ref{fig:average_longime} shows the horizontally averaged temperature at $z = 10$~Mm as a function of time. A Gaussian filter was applied to the profiles to smooth out short-lived fluctuations and highlight overall trends. Across all three BC implementations, the rate of thermal decay is strictly domain-height dependent: shorter domains ($z_{\text{max}} \le 26.2\text{ Mm}$) cool significantly faster and reach much lower temperatures than taller domains ($z_{\text{max}} \ge 34.3\text{ Mm}$). However, the choice of upper BC fundamentally dictates the speed, continuity, and late-time stability of this cooling process.

The top panel (dynamic top BC), shows rapid early cooling, particularly for the three shortest domain heights ($14.1\text{ Mm}$, $18.2\text{ Mm}$, and $26.2\text{ Mm}$). These short domains experience a steep drop within the first $0.4\text{ h}$ to $0.7\text{ h}$ before leveling off into distinct, quasi-steady thermal plateaus between $6 \times 10^4\text{ K}$ and $10^5\text{ K}$. Taller domains undergo much gentler relaxation, maintaining temperatures above $3 \times 10^5\text{ K}$ throughout the run. Notably, the stretched-grid simulation is almost identical to its uniform-grid counterpart, confirming numerical consistency.

The middle panel (closed top BC) displays a smoother, more delayed cooling trend across all domain heights. Compared to the dynamic top BC runs, energy loss occurs at a reduced rate; only the shortest domain ($14.1\text{ Mm}$) manages to reach the $\sim 8 \times 10^4\text{ K}$ thermal floor, doing so much later at $t \approx 0.9\text{ h}$. Intermediate heights decay at a steady, nearly constant rate on the logarithmic scale, while taller domains remain elevated above $4 \times 10^5\text{ K}$ with minimal overall temperature drop over the $1.1\text{ h}$ duration.

The bottom panel (non-reflecting characteristic top BC) demonstrates the strongest thermal retention during early and mid-stage evolution. Across nearly all domain heights, temperatures remain largely insulated and constant near $7 \times 10^5\text{ K}$ to $9 \times 10^5\text{ K}$ up to $t \approx 0.8\text{ h}$. Beyond $t = 0.8\text{ h}$, the shortest runs ($14.1\text{ Mm}$, $18.2\text{ Mm}$, and $26.2\text{ Mm}$) experience sharp, late-time collapses, dropping steeply down toward $1.5 \times 10^5\text{ K}$ to $4 \times 10^5\text{ K}$, whereas taller domains maintain high, virtually unchanged thermal profiles throughout the run.

\begin{figure*}
\centering
\includegraphics[width=0.8\linewidth]{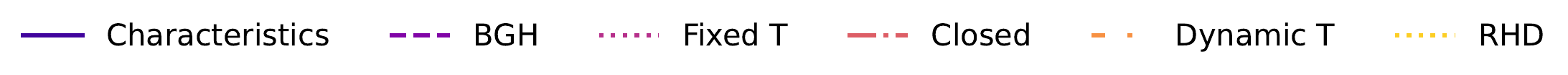}\\
\includegraphics[width=0.4\linewidth]{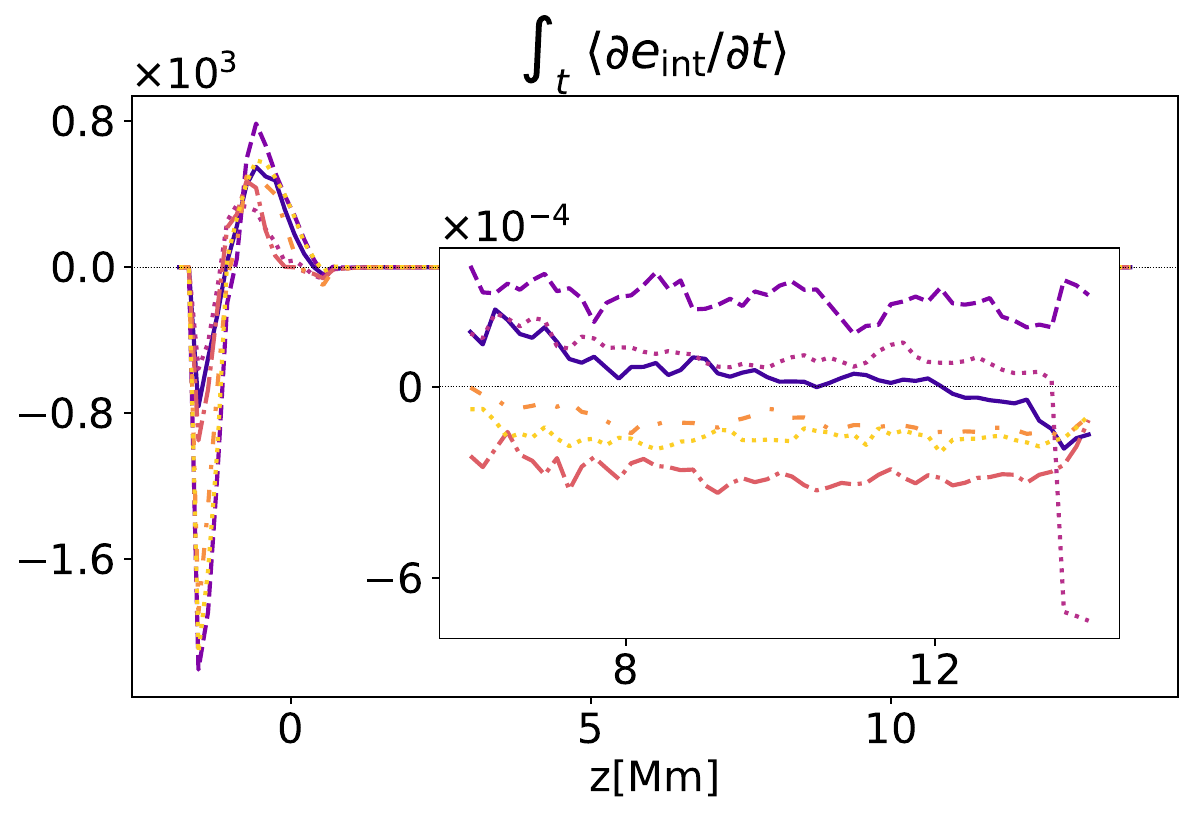 }
\includegraphics[width=0.4\linewidth]{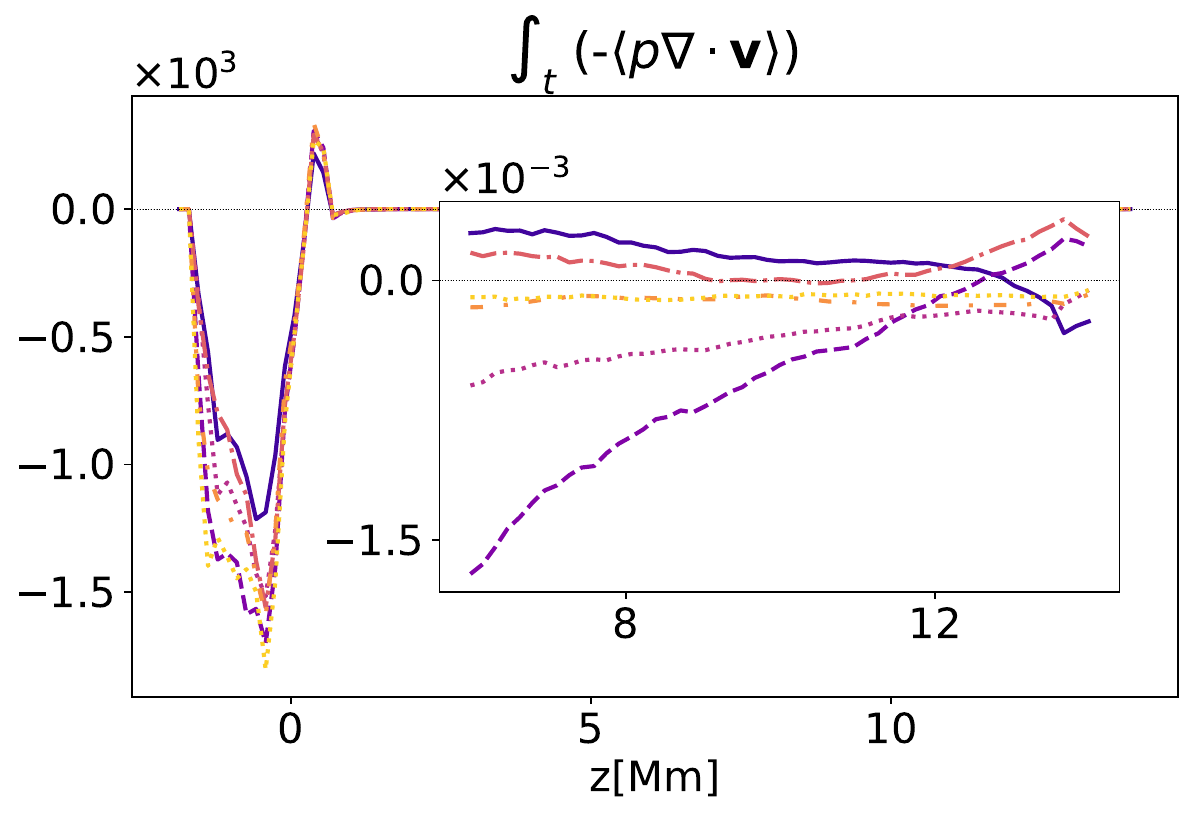}\\ 
\includegraphics[width=0.4\linewidth]{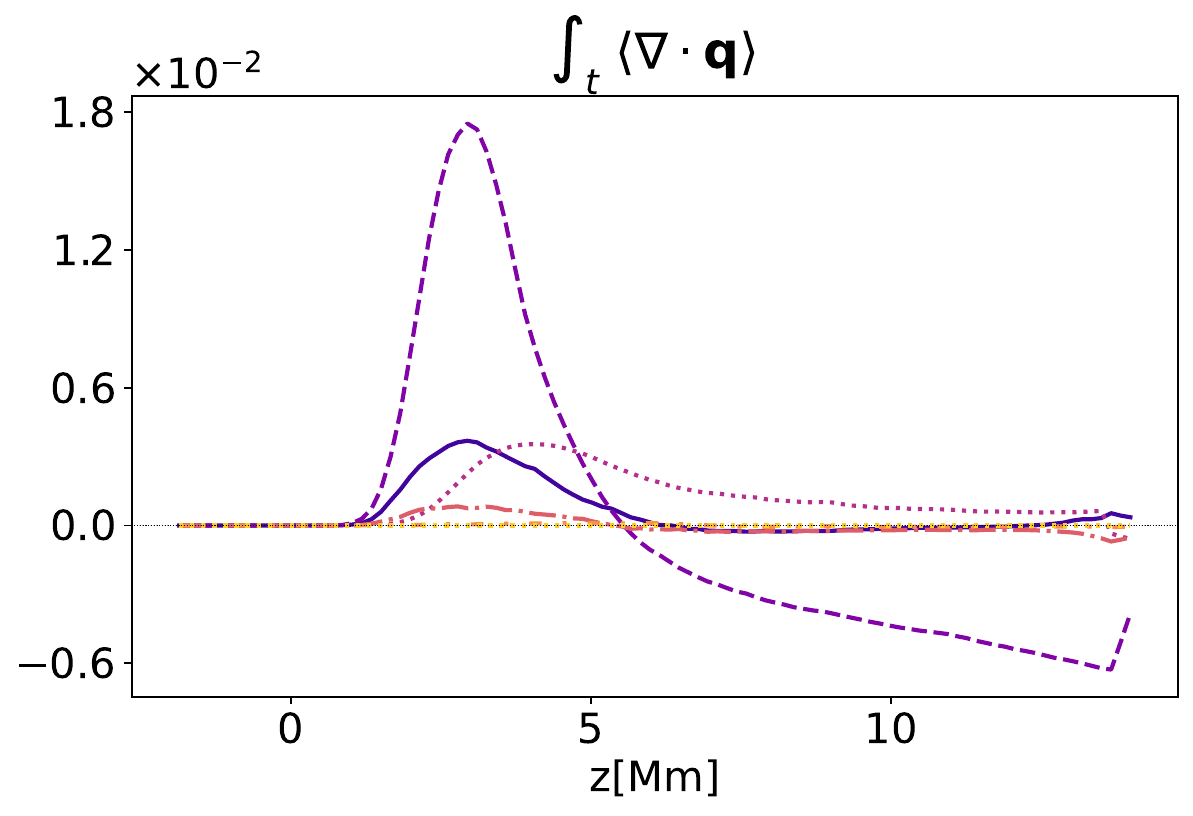} 
\includegraphics[width=0.4\linewidth]{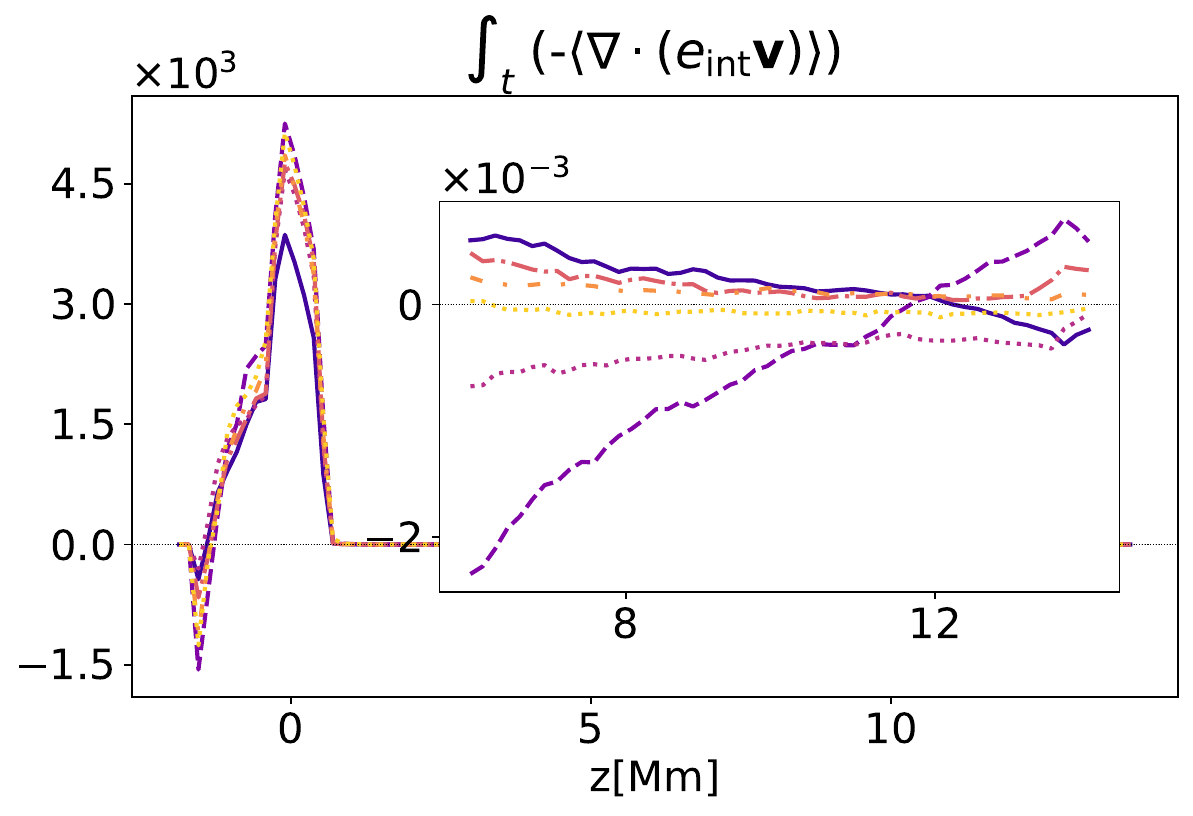}\\
\includegraphics[width=0.4\linewidth]{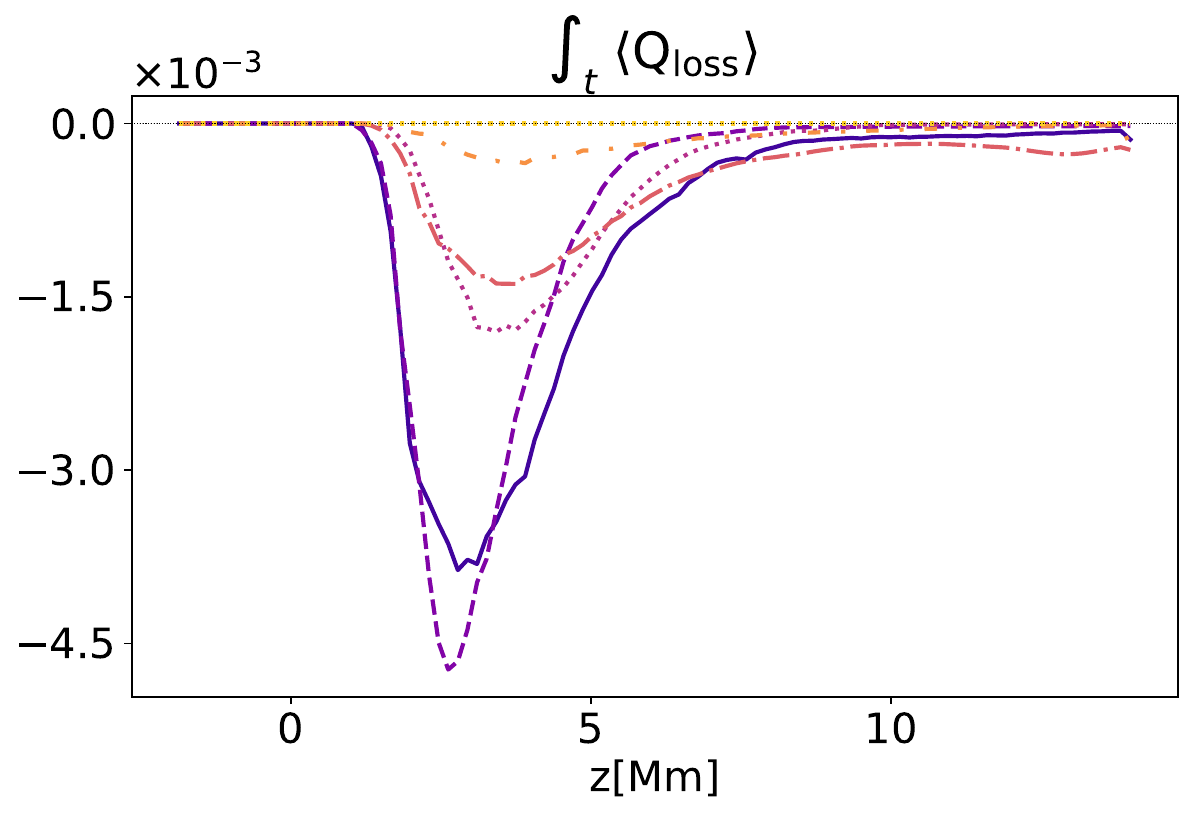}
\includegraphics[width=0.4\linewidth]{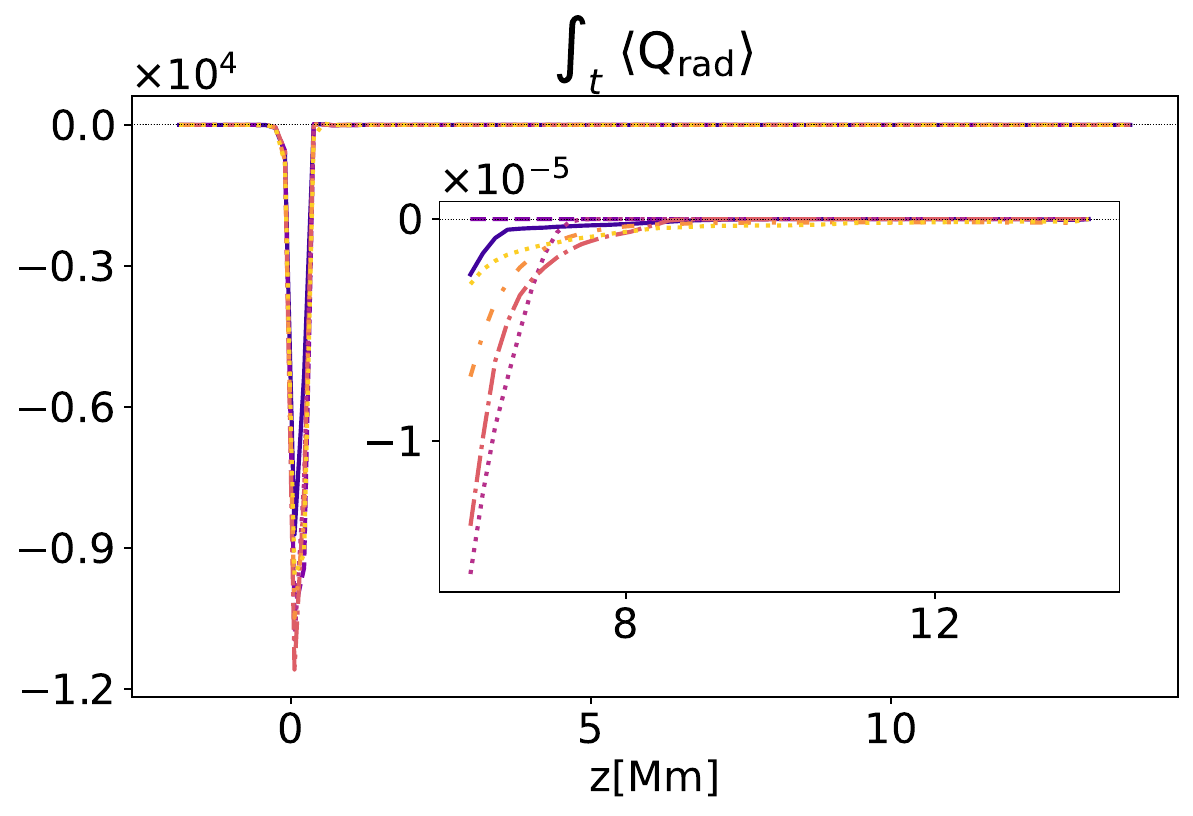}
\caption{Horizontally averaged energetic contributions from the internal energy equation \eqref{eq:evol_eint}, integrated over time and plotted as a function of height. The analysis is restricted to simulations with $z_{\max} = 14.1~\mathrm{Mm}$ and covers a 30-minute interval ($t = 1200\text{--}3000~\mathrm{s}$). Each panel displays a different energetic term, comparing the BGH model (set~J), the fixed top BC (set~D), the dynamic top BC (set~E), the closed BC (set~H), the RHD model (set~C), and the model with characteristic BC. The horizontal black lines indicate the zero level. The units are in kg m$^{-1}$s$^{-2}$.}
\label{fig:heating_cooling}
\end{figure*}

\subsection{ Analysis of energy contributions \label{subsec:energetic_analysis} }

In this section, we aim to understand the differences in the obtained temperature profiles in the different sets of simulations by exploring and comparing the individual energetic contributions from each term in the internal energy equation. Consider the evolution equation for internal energy:
\begin{eqnarray}
& & \frac{\partial {\rm e}_{\rm int}}{\partial t} = -\nabla\cdot\left({\rm e}_{\rm int} \bf{v}\right) -  p\nabla \cdot {\bf v}  - \nabla \cdot \bf{q} +  \qlossEQ + \qradEQ  , \label{eq:evol_eint}
\end{eqnarray}
where ${\rm e}_{\rm int}$ is the internal energy, $\bf{v}$ is the velocity, $p$ is the pressure, $\bf{q}$ is the thermal conduction vector, $\qlossEQ$ represents the optically thin radiative losses, and $\qradEQ$ corresponds to the optically thick radiative term. The terms in the equation are; $-\nabla\cdot\left({\rm e}_{\rm int} \bf{v}\right)$: internal energy flux, -$p\nabla \cdot {\bf v}$: pressure force work, and -$\nabla \cdot \bf{q}$: thermal conduction contribution.  

In Fig.~\ref{fig:heating_cooling}, we present the horizontally averaged terms of the internal energy equation (Eq. \ref{eq:evol_eint}), retaining its algebraic sign and integrated over time and plotted as a function of height. For this comparison, we focus exclusively on the simulations with $z_{\max} = 14.1~\mathrm{Mm}$. The analysis spans a 30-minute interval from $t = 1200\text{--}3000~\mathrm{s}$, matching the time period used for the analysis in Sec. 3.1 to 3.3. Each panel displays a specific energetic term across the different simulation types from all sets: the non-reflecting characteristic BC run (set~L), the BGH run (set~J), the fixed top BC (set~D), the dynamic top BC (set~E), the closed BC (set~H), and the RHD run (set~C). For reference purposes a horizontal black line marks the zero level and for a clearer view, some panels also include a zoomed-in inset showing the corona from $z = 6~\mathrm{Mm}$ upward.

In the deep atmospheric layers ($z \lesssim 1~\mathrm{Mm}$), the energy budget is heavily dominated by localized radiative transport. The primary energy sink is the optically thick radiative cooling term, \qradint{}, which shows a sharp, negative delta-like spike localized precisely at the optical surface ($z \approx 0~\mathrm{Mm}$). Below the surface, the pressure work term (\peworkint{}) contributes a large cooling sink. Above the surface, it transitions to a positive heating contribution, reaching a peak of approximately $0.5\times 10^3$.  Notably, all six models display nearly identical profiles in this deep zone. This convergence demonstrates that the choice of the upper BC (or the absence of coronal energy terms, as in the RHD simulation) does not pollute the lower-atmospheric dynamics within the simulated time-frame.

As the plasma density drops with height, the atmosphere transitions into a regime where optically thin radiative losses, \qlossint{}, become efficient. As seen in the left panel of the third row, all magnetic models present a distinct cooling valley between $z \sim 2~\mathrm{Mm}$ and $\sim 5~\mathrm{Mm}$, tracking the location of the average TR. The BGH model and the one with non-reflecting characteristic BC exhibit the most pronounced radiative losses in this layer. The primary balancing agent for this thin radiative loss varies strictly by model. In the BGH case, the energy deficit is strongly compensated by the thermal conduction flux divergence, \divqtcint{}, which shows a prominent positive heating peak centered at $z \sim 3~\mathrm{Mm}$.
On the other hand, the alternative setups exhibit more suppressed thermal conduction profiles through this layer, though this is the zone where their accumulated values are strongest. It is interesting to notice that they have positive values at almost all heights in the fixed boundary simulation (hot-plate case), whereas for the other simulations it is mainly negative in the corona.

For $z >$ 5 Mm, the energetic terms drop by several orders of magnitude, as highlighted by the inset plots. In this regime, the divergence between the  BGH model and the others becomes critical. Regarding the net internal energy rate \rhseintint{}, which is calculated as the sum of all the other terms, we found that the BGH model sustains a net positive internal energy accumulation throughout the corona, as well as most of the corona for the model with fixed BC temperature (hot-plate case). For the model with non-reflecting characteristic BC, the net sum remains positive below $12\text{ Mm}$—though lower in magnitude than in both the BGH and fixed-BC setups—before shifting to negative values above $12\text{ Mm}$. The other runs have values below zero, explaining why only the BGH and the fixed-BC model are able to sustain a stable hot 1MK corona.

Regarding the pressure work, \peworkint{}, above 5 Mm, the run with non-reflecting characteristic BC exhibits a sign change from positive to negative at around 13 Mm. Conversely, the BGH model transitions from negative to positive at a similar height. In the closed BC simulation, this term remains positive throughout the corona, which we attribute to continuous wave reflections bouncing vertically off the upper boundary. Across all other simulations, pressure work consistently acts as a cooling agent throughout the corona. The advected internal energy, \eintfluxint{}, shows behavior similar to \peworkint{}, with the exception of the dynamic top BC run, where it maintains positive values in the corona.

We note that while the present analysis accounts for each explicit term in the internal energy equation, a complete energy budget must also incorporate the implicit contributions from numerical dissipation. A detailed treatment of these implicit effects is deferred to future work.

\section{Conclusions\label{sec:conclusions}}

In this work, we conducted an extensive study of the combined effects of computational domain size and temperature treatment at the top BC in two-dimensional numerical simulations of the solar atmosphere in open-field regions. We found that variations in the horizontal extent of the domain, ranging from 5.8~Mm to 46.1~Mm, have no significant impact on the average thermal profile.

In general, simulations with larger vertical extensions sustained higher average temperatures at intermediate coronal heights (above $\sim 10\text{ Mm}$). Additionally, the horizontally averaged profiles of the taller boxes exhibit a steeper temperature gradient in the average TR.

The RHD simulations have some of the sharpest average TR among all the experiments considered. This behavior is likely related to the absence of thermal conduction and coronal radiative losses in these models, which allows the average TR structure of the initial atmosphere to be largely preserved and continuously reinforced by convective motions and shock propagation. Models with larger vertical domains sustain higher coronal temperatures initially; however, these states are not in thermal equilibrium and gradually cool over time, with the cooling rate being lower in simulations with greater box heights.

RMHD simulations with a hot-plate BC show a distinct height dependence, where taller domains maintain higher mean temperatures in the lower corona and a more vertically confined average TR profile. This behavior is observed only for box heights exceeding 26.1~Mm, suggesting that simulations using a hot-plate with smaller vertical domains may not be advisable because in those cases we can see a direct influence of the boundary in the physical profile. RMHD simulations with a dynamic top BC show a similar height dependence. For box heights larger than 30~Mm, the average atmospheric profiles become more stable in time, with both coronal temperature and TR sharpness increasing. Our results show that only sufficiently large domains (z$_\text{max}$ $\ge$ 46 Mm), combined with dynamic BC, can yield an atmosphere below $z = 10$ Mm that differs by less than 24\% from that obtained with the hot-plate condition.

The simulations using a closed BC show profiles that fall between the fixed-temperature (hot-plate) runs and those with a dynamic top. While these closed-boundary simulations can maintain a hotter corona compared to the dynamic case, they are still unable to sustain it over longer timescales. Simulations with non-reflecting characteristic BC show weaker height dependence, producing average profiles with much less variation across box heights than observed in other configurations. Nevertheless, taller domains in characteristic setups still support higher coronal temperatures over longer timescales. Their stability significantly outperforms both closed and dynamic BC setups, maintaining steady average temperatures at 10 Mm above $8\times 10^5~\mathrm{K}$ for more than 1 solar hour. The BGH models exhibit a remarkably steep average TR that is downward-shifted by $1~\mathrm{Mm}$ relative to the other setups, a direct consequence of the explicitly imposed heating. These models sustain the highest coronal temperatures, which further increase with larger domain heights. 

For simulations utilizing a stretched grid and a dynamic top BC, the results were highly consistent. Across these runs, we observed a reduction of the computational cost while preserving the average atmospheric profiles. The fastest stretched-grid simulation achieved a speed-up factor of 2.8. We found that a stretching factor of \(S = 3.7\), with grid spacing increasing from 14~km to approximately 193~km up to \(z = 20\)~Mm and reaching about 800~km at \(z = 54.4\)~Mm, provides an effective compromise between accuracy and computational efficiency. Moreover, for simulations of identical height, the stretched-grid runs cool down more slowly in the corona than the uniform-grid runs, reaching the asymptotic temperature roughly one hour later.  From this set of experiments, we conclude that the most critical parameters are the vertical extent of the computational domain and the degree of grid stretching. In addition, we gain insight into the timescale over which this type of simulation can maintain coronal temperatures above 10$^5$ K, which varies between 30 minutes and 3 hours as a function of the domain height.

The energetic analysis revealed that while the deep atmospheric layers remain robustly independent of upper BC and coronal terms (heat conduction and coronal radiative losses), the energetic budget of the upper corona is critically sensitive to them, shifting the balance between different terms.

Only the BGH and fixed-temperature (hot-plate) models sustain a net positive internal energy rate, \rhseintint{}, throughout the corona, directly explaining their unique ability to maintain a stable, 1 MK coronal temperature. Nevertheless, among the remaining models, those employing non-reflecting BC delay coronal cooling more effectively and maintain relatively high temperatures for at least one solar hour, provided the domain height exceeds 34 Mm. This demonstrates that the energetic balance in the corona is highly complex and depends heavily on the treatment of the upper boundary.

Although strictly valid for 2D configurations, these results provide vital foundational benchmarks for state-of-the-art modeling. The computational economy of 2D setups enabled an exhaustive parameter space exploration that would be prohibitively expensive in 3D. Through this approach, we systematically evaluated and compared the most prevalent modeling strategies used by the community, including configurations with a BGH, closed boundaries, dynamic top boundaries that permit outward flows, and traditional hot-plate. Ultimately, we mapped out how the spatial design of the numerical domain and the choice of boundary constraints directly dictate the averaged thermal profile and long-term stability of the simulated atmosphere, offering a calibrated guide for setting up more complex, multi-dimensional simulations.

Our results confirm the necessity of 3D simulations, which are able to sustain a hot corona, likely via magnetic field braiding \citep{2005ApJ...618.1020G, 2017ApJ...834...10R,2022ApJ...937...91C}. Simple 2D vertical magnetic fields fail to maintain these temperatures without auxiliary heating or artificial adjustments. This highlights the critical need for either explicit 3D dynamics or the coupling of regions with distinct magnetic topologies, such as interacting active and plage areas \citep[e.g.,][]{2017ApJ...847...36M}.

\begin{acknowledgements}
This research was funded by the Agencia Estatal de Investigación del Ministerio de Ciencia, Innovación y Universidades (MCIU/AEI) and the European Regional Development Fund (ERDF) under grants "MHD MODeling and beyond: in preparation for the European Solar Telescope" (PID2021-127487NB-I00) and "Solar atmosphere dynamics: connectivity, energy transfer, and diagnostics" (PID2024-156538NB-I00), and by the MCIU/AEI and the European Social Fund (ESF) under grant "Ayudas Ramón y Cajal Convocatoria 2020" with reference RYC2020-030307-I. TF acknowledges support from the Project CNS2023-145233, funded by MICIU/AEI/10.13039/501100011033, and the European Union “NextGenerationEU/RTRP”. The authors gratefully acknowledge the technical support and expertise provided by the Spanish Supercomputing Network (Red Española de Supercomputación). Computational resources were provided by the LaPalma Supercomputer at the Instituto de Astrofísica de Canarias and the MareNostrum Supercomputer in Barcelona, Spain. This work is also part of grant CEX2025-001609-S, awarded to the Instituto de Astrofísica de Canarias under the Severo Ochoa Centre of Excellence program and funded by MICIU/AEI/10.13039/501100011033.
\end{acknowledgements}


\begin{thebibliography}{21}
\expandafter\ifx\csname natexlab\endcsname\relax\def\natexlab#1{#1}\fi

\bibitem[{{Carlsson} {et~al.}(2016){Carlsson}, {Hansteen}, {Gudiksen}, {Leenaarts}, \& {De Pontieu}}]{2016A&A...585A...4C}
{Carlsson}, M., {Hansteen}, V.~H., {Gudiksen}, B.~V., {Leenaarts}, J., \& {De Pontieu}, B. 2016, \aap, 585, A4

\bibitem[{{Chen} {et~al.}(2022){Chen}, {Rempel}, \& {Fan}}]{2022ApJ...937...91C}
{Chen}, F., {Rempel}, M., \& {Fan}, Y. 2022, \apj, 937, 91

\bibitem[{{Evans} \& {Hawley}(1988)}]{CT_evans}
{Evans}, C.~R. \& {Hawley}, J.~F. 1988, \apj, 332, 659

\bibitem[{Gottlieb {et~al.}(2001)Gottlieb, Shu, \& Tadmor}]{gottlieb2001ssp}
Gottlieb, S., Shu, C.-W., \& Tadmor, E. 2001, SIAM Rev., 43, 89

\bibitem[{{Gudiksen} \& {Nordlund}(2005)}]{2005ApJ...618.1020G}
{Gudiksen}, B.~V. \& {Nordlund}, {\r{A}}. 2005, \apj, 618, 1020

\bibitem[{{Heggland} {et~al.}(2011){Heggland}, {Hansteen}, {De Pontieu}, \& {Carlsson}}]{2011ApJ...743..142H}
{Heggland}, L., {Hansteen}, V.~H., {De Pontieu}, B., \& {Carlsson}, M. 2011, \apj, 743, 142

\bibitem[{{Iijima} \& {Yokoyama}(2015)}]{2015ApJ...812L..30I}
{Iijima}, H. \& {Yokoyama}, T. 2015, \apjl, 812, L30

\bibitem[{{Khomenko} {et~al.}(2018){Khomenko}, {Vitas}, {Collados}, \& {de Vicente}}]{2018A&A...618A..87K}
{Khomenko}, E., {Vitas}, N., {Collados}, M., \& {de Vicente}, A. 2018, \aap, 618, A87

\bibitem[{{Leenaarts} {et~al.}(2011){Leenaarts}, {Carlsson}, {Hansteen}, \& {Gudiksen}}]{2011A&A...530A.124L}
{Leenaarts}, J., {Carlsson}, M., {Hansteen}, V., \& {Gudiksen}, B.~V. 2011, \aap, 530, A124

\bibitem[{{Li}(2005)}]{hllc_mhd_Li}
{Li}, S. 2005, J. Comput. Phys., 203, 344

\bibitem[{Makaremi-Esfarjani \& Najafi-Yazdi(2022)}]{MAKAREMIESFARJANI2022105461}
Makaremi-Esfarjani, P. \& Najafi-Yazdi, A. 2022, Comput. Fluids, 241, 105461

\bibitem[{{Mart{\'\i}nez-Sykora} {et~al.}(2017){Mart{\'\i}nez-Sykora}, {De Pontieu}, {Carlsson}, {Hansteen}, {N{\'o}brega-Siverio}, \& {Gudiksen}}]{2017ApJ...847...36M}
{Mart{\'\i}nez-Sykora}, J., {De Pontieu}, B., {Carlsson}, M., {et~al.} 2017, \apj, 847, 36

\bibitem[{{Mignone} {et~al.}(2007){Mignone}, {Bodo}, {Massaglia}, {Matsakos}, {Tesileanu}, {Zanni}, \& {Ferrari}}]{2007ApJS..170..228M}
{Mignone}, A., {Bodo}, G., {Massaglia}, S., {et~al.} 2007, \apjs, 170, 228

\bibitem[{{Navarro} {et~al.}(2025){Navarro}, {Khomenko}, {Vitas}, \& {Felipe}}]{magec}
{Navarro}, A., {Khomenko}, E., {Vitas}, N., \& {Felipe}, T. 2025, \aap, 703, A298

\bibitem[{{N{\'o}brega-Siverio} {et~al.}(2016){N{\'o}brega-Siverio}, {Moreno-Insertis}, \& {Mart{\'\i}nez-Sykora}}]{2016ApJ...822...18N}
{N{\'o}brega-Siverio}, D., {Moreno-Insertis}, F., \& {Mart{\'\i}nez-Sykora}, J. 2016, \apj, 822, 18

\bibitem[{{Rempel}(2017)}]{2017ApJ...834...10R}
{Rempel}, M. 2017, \apj, 834, 10

\bibitem[{{Roe}(1986)}]{Roe1986}
{Roe}, P.~L. 1986, Annu. Rev. Fluid Mech., 18, 337

\bibitem[{Roe \& Balsara(1996)}]{roe_balsara_eigenvectors}
Roe, P.~L. \& Balsara, D.~S. 1996, SIAM J. Appl. Math., 56, 57

\bibitem[{{Sen} \& {Keppens}(2022)}]{2022A&A...666A..28S}
{Sen}, S. \& {Keppens}, R. 2022, \aap, 666, A28

\bibitem[{{Spruit}(1974)}]{1974SoPh...34..277S}
{Spruit}, H.~C. 1974, \solphys, 34, 277

\bibitem[{{Vernazza} {et~al.}(1981){Vernazza}, {Avrett}, \& {Loeser}}]{Vernazza_temperature_profile}
{Vernazza}, J.~E., {Avrett}, E.~H., \& {Loeser}, R. 1981, \apjs, 45, 635

\end{thebibliography}
\end{document}